\documentclass[%
superscriptaddress,
amsmath,amssymb,amsfonts,
twocolumn,
aps,
]{revtex4-2}

\pdfoutput=1 
\usepackage{graphicx}
\usepackage{bm} 
\usepackage{physics}
\usepackage{amsthm}
\usepackage{amsmath}
\usepackage{xcolor}
\usepackage{textcomp}
\usepackage{outlines}
\usepackage{soul}
\usepackage{bm}
\usepackage[colorlinks=true, allcolors=black]{hyperref}
\usepackage[toc,page]{appendix}
\usepackage{cleveref}
\usepackage{multirow}

\usepackage{float}
\usepackage{mdframed}    

\newcommand{\dual}{\mathcal{D}}

\newcommand{\G}{\mathbb{G}}

\newcommand{\Lag}{\mathcal{L}}

\newcommand{\V}{\mathbb{V}}

\newcommand{\mI}{\vb{I}}

\newcommand{\vei}{\bm e_i}
\newcommand{\vo}{\bm o}
\newcommand{\mO}{\bm O}
\newcommand{\vp}{\bm p}
\newcommand{\vv}{\bm v}
\newcommand{\vx}{\bm x}
\newcommand{\vz}{\bm z}

\newcommand{\vgma}{\bm \gamma}
\newcommand{\vphi}{\bm \phi}
\newcommand{\mGv}{\bm G_0}
\newcommand{\mU}{\bm U}
\newcommand{\mZ}{\bm Z}

\newcommand{\Diag}{\text{Diag}}

\newcommand\numthis{\stepcounter{equation}\tag{\theequation}}
\usepackage{mathtools}

\renewcommand{\eqref}[1]{Eq.~(\ref{#1})}

\newcommand{\REVISION}[1]{{\color{red} #1}}
\DeclareUnicodeCharacter{202F}{\REVISION{There is a 202F unicode character here}}

\begin{document}

\preprint{APS/123-QED}

\title{Bounds as blueprints: towards optimal and accelerated photonic inverse design}
 \author{Pengning~Chao\(^\ddagger\)}
 \email{Contact author: pchao827@mit.edu}
 \affiliation{Department of Mathematics, Massachusetts Institute of Technology, Cambridge, Massachusetts 02139, USA}
 \thanks{These authors contributed equally to this work.}
 \author{Alessio~Amaolo\(^\ddagger\)}
 \email{Contact author: alessioamaolo@princeton.edu}
 \affiliation{Department of Chemistry, Princeton University, Princeton, New Jersey 08544, USA}
 \thanks{These authors contributed equally to this work.}
 \author{Sean~Molesky}
 \affiliation{Department of Engineering Physics, Polytechnique Montréal, Montréal, Québec H3T 1J4, Canada}
 \author{Alejandro~W.~Rodriguez}
 \affiliation{Department of Electrical and Computer Engineering, Princeton University, Princeton, New Jersey 08544, USA}
\date{\today}   

\begin{abstract}
Our ability to structure materials at the nanoscale has, and continues to, enable key advances in optical control. In pursuit of optimal photonic designs, substantial progress has been made on two complementary fronts: bottom-up structural optimizations (inverse design) discover complex high-performing structures but offer no guarantees of optimality; top-down field optimizations (convex relaxations) reveal fundamental performance limits but offer no guarantees that structures meeting the limits exist. We bridge the gap between these two parallel paradigms by introducing a ``verlan'' initialization method that exploits the encoded local and global wave information in duality-based convex relaxations to guide inverse design towards better-performing structures. We illustrate this technique via the challenging problem of Purcell enhancement, maximizing the power extracted from a small emitter in the vicinity of a photonic structure, where ill-conditioning and the presence of competing local maxima lead to sub-optimal designs for adjoint optimization. Structures discovered by our verlan method outperform standard (random) initializations by close to an order of magnitude and approach fundamental performance limits within a factor of two, highlighting the possibility of accessing significant untapped performance improvements. 
\end{abstract}
\maketitle

\section{Introduction}

\begin{figure*}[]
    \centering
    \includegraphics[width=\linewidth]{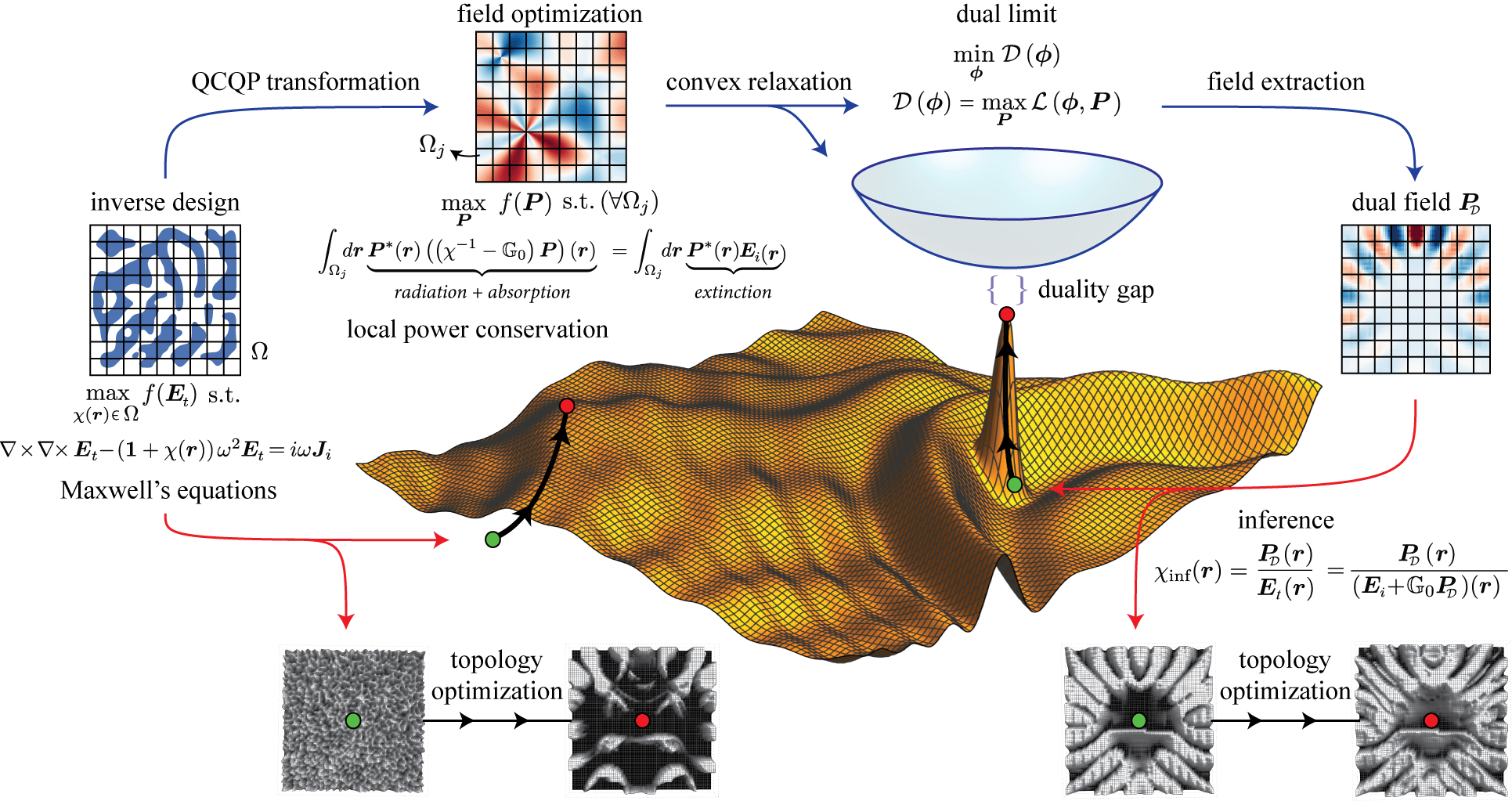}
    \caption{\textbf{Schematic of verlan method}. The  optimization landscape of photonic design---here illustrated as a surface, but in practice having thousands of dimensions---is often very complicated, exhibiting ill-conditioning and a large number of local optima which can trap gradient-based (adjoint) algorithms. Limits to device performance can nevertheless be evaluated by transforming the inverse design problem into a quadratically constrained quadratic field optimization program (QCQP), and then applying a convex relaxation (e.g. Lagrange duality) to create a dual limit program. In contrast to structural optimization, the dual program (minimization over Lagrange multipliers) is convex, with a unique global minimum corresponding to the tightest bound. The ``polarization'' associated with the dual optimum contains global information on the characteristics of optimal fields. 
    Retracing this outer ring of transformations illustrates the major steps involved in our verlan method.  
    First, the original inverse design problem is converted into a QCQP over the polarization field. 
    Next, this QCQP is convexified via Lagrange duality to produce a dual program, which determines a super-optimal dual polarization field $\bm{P}_{\dual}$.
    Using \eqref{eq:inference}, a material profile is then inferred from $\bm{P}_{\dual}$ and used to initialize a local optimization (e.g. topology optimization).}
    \label{fig:schematic}
\end{figure*}

The design of optical components for technological applications has relied for many decades on composing a small number of well-understood physical mechanisms for light control, e.g. index guiding~\cite{yariv_photonics_2006}, photonic bandgaps~\cite{joannopoulos_photonic_2008}, and material resonances~\cite{maier_plasmonics_2007}. However, with growing freedom for structuring materials down to the nanoscale (even in three dimensions~\cite{roberts_3d_2023}), and increasing demands on device functionality and complexity, this paradigm is currently being partially supplanted by computational inverse design---making minimal assumptions on structure geometry or operating principles, the act of design is equated with an optimization over an enormous number of parameterizing degrees of freedom to discover effective devices (left side Fig.~\ref{fig:schematic})~\cite{jensenTopologyOptimization2011,molesky_inverse_2018,liEmpoweringMetasurfaces2022}.
Such large-scale optimization techniques have found remarkable success in myriad domains~\cite{lalau-keraly_adjoint_2013,hughes_adjoint_2018,piggottInverseDesignedPhotonics2020,hammondPhotonicTopology2021,chenInverseDesign2022,stich_inverse_2024,roques-carmes_metaoptic_2025}, inspiring photonics engineers to push towards ever higher performances and tighter form factors. 

At the frontier of our engineering capability, the high-dimensional and nonconvex nature of the inverse design problem often leads to uncertainty (center Fig.~\ref{fig:schematic}): adjoint gradient-based algorithms are generally necessary for navigating the high-dimensional design space, and if an optimization fails to achieve a desired level of performance, it may be that the algorithm is simply ``stuck" in a sub-par local optimum, or it may be that the desired performance is physically impossible to achieve. To address this uncertainty, recent progress has been made on elucidating fundamental limits to photonic performance which no structure may exceed (e.g. duality limits, top right Fig.~\ref{fig:schematic}). The key to such performance bounds rests on a reformulation of the structural design problem to an optimization over fields (instead of geometry) subject to power conservation constraints that are quadratic functions of the fields (top left Fig.~\ref{fig:schematic})~\cite{angerisHeuristicMethods2021,chaoPhysicalLimits2022}. When the design objective is also a quadratic  function of the fields, this leads to a quadratically constrained quadratic program (QCQP)~\cite{parkGeneralHeuristics2017} which can be bounded by convex relaxation techniques such as Lagrangian duality~\cite{parkGeneralHeuristics2017,chaoPhysicalLimits2022} or equivalently, lifting to a semi-definite program (SDP)~\cite{angerisHeuristicMethods2021,gertlerManyPhotonic2025}. This recipe has been applied to many important problems~\cite{gustafssonUpperBounds2020,molesky_global_2020,kuang_maximal_2020,molesky_hierarchical_2020,kuang_computational_2020,jelinekFundamentalBounds2021,zhangConservationlawbasedGlobal2021,molesky_t_communication_2022,chaoMaximumElectromagnetic2022,angerisBoundsEfficiency2023,amaolo_can_heterostructures_2023,mohajanFundamentalLimits2023,strekhaTraceExpressions2024,amaoloPhysicalLimits2024,amaoloMaximumShannon2024,chaoSumofSquaresBounds2025}, producing limits that often closely approach the performance achieved by inverse design. When this happens, the resulting bounds certify the optimality of known designs~\cite{venkataram_fundamental_2020-1}, and can be used to deduce scaling laws for characteristic engineering parameters such as device size~\cite{liThicknessBound2022,liHighefficiencyHighnumericalaperture2024} and material choice~\cite{shim_fundamental_2019,chao_maximum_2022}. 

When there is a significant difference between bounds and existing designs, the situation is intriguing in another way: are the limits loose, signifying a large duality gap, or are there as-yet undiscovered superior designs? The second possibility deserves serious consideration: the wave nature of light allows for complicated interference patterns that create a large number of local structural optima (center Fig.~\ref{fig:schematic}). Moreover, photonic devices often rely on physical effects that are highly sensitive to small structural perturbations (e.g., high-quality resonances~\cite{armani_ultra-high-q_2003,hsuBoundStates2016}, multipole cancellation~\cite{johnsonMultipolecancellationMechanism2001,srinivasanMomentumSpace2002}, localized fields~\cite{maierPlasmonicsLocalization2005,albrechtsenNanometerscalePhoton2022}), making structural optimization numerically ill-conditioned~\cite{liangFormulationScalable2013}. 
Photonic inverse design with adjoint gradients is thus strongly contingent on the starting point~\cite{chenValidationCharacterization2024}, and given the high dimensional design space, it is impossible to exhaustively sample different initializations. A natural question is whether the limits can provide information on optimal device characteristics that can point inverse design in the right direction. 

The few existing preliminary explorations of this line of inquiry have combined the SDP formulation of performance limits with low-rank approximations of the SDP matrix variable to obtain physically feasible designs. To date, these ideas have only been applied to small-scale problems (1D multilayer reflectors~\cite{gertlerManyPhotonic2025} and 2D sub-wavelength mode converters~\cite{dalklintPerformanceBounds2024}), and the resulting structures did not improve upon the state of the art, performing at or below the level of best inverse designs. 

In this paper, we report a novel method using Lagrangian dual limits to extract structure templates for guiding and accelerating inverse design. 
Throughout the text, to differentiate from current initialization methods based primarily on random sampling (hereafter referred to as ``standard'') and emphasize the role of duality, we refer to our new approach as ``verlan''---the syllabic inversion of ``l'enverse'' (the reverse) in verlan, a french language argot. 
As a representative example, we apply verlan extraction with topology optimization (TopOpt) to discover 2D structures that maximize the Purcell enhancement of nearby out-of-plane point sources, an ill-conditioned structural optimization problem that is a challenging benchmark for existing inverse design approaches~\cite{chenValidationCharacterization2024}. The method is numerically efficient, enabling us to evaluate bounds and extract structures for large square design regions up to six vacuum wavelengths along each side, involving (sparse) system matrices with size up to $57600\times57600$. Near this maximal size, verlan inverse designs achieved over one thousand-fold Purcell enhancement, in spite of the choice of a low-index and lossy material, outperforming TopOpt with standard initializations by nearly an order of magnitude and approaching the performance limits to within a factor of two. Verlan designs also exhibit qualitatively different geometric features, exploiting distinct and hitherto unexplored enhancement mechanisms. Our results represent a significant step towards a practical general-purpose framework for integrating the global information encoded in limits into photonic inverse design. 

While the basic motivation of our approach is entirely intuitive, it is equally reinforced by an analysis of Sion's minimax theorem~\cite{sionGeneralMinimax1958} as applied to field optimization QCQPs. 
This analysis also underpins certain technical aspects mentioned here only in passing.  
Readers interested in these details (e.g. scraping) may consult the partner manuscript~\cite{molesky_verlan} for additional explanation.

\section{Methods}

Figure~\ref{fig:schematic} depicts a schematic overview of inverse design, Lagrangian dual limits, and the proposed verlan methodology. For simplicity, although much greater generality is possible~\cite{chao_physical_2022,mohajan_fundamental_2023}, we restrict our attention to linear isotropic media in the frequency domain, with $\omega=2\pi/\lambda$ denoting the angular frequency and $\lambda$ the vacuum wavelength. The photonic device to be optimized is defined by a binary susceptibility profile $\chi(\bm r) \in \{0,\chi\}$ for some constant complex electric susceptibility $\chi$. Following the general framework of structural optimization, we consider maximizing any desired objective function $f$ of the electromagnetic fields (e.g., Purcell enhancement, focusing efficiency, cross sections) by tuning $\chi(\bm r)$ within a specified design region $\Omega$:
\begin{align*}
    \max_{\chi(\bm r)\in \Omega} \quad & f(\bm E_t) \numthis\label{eq:topopt} \\ 
    \textrm{s.t.} \quad & \nabla \times \nabla \times \bm E_t - (\bm 1+\chi(\bm r)) \omega^2 \bm E_t = i\omega \bm J_i, \\
    & \chi(\bm r) =  0 \text{ or } \chi,
\end{align*}
where $\bm J_i$ is a source current driving the total electric field $\bm E_t$ in dimensionless electromagnetic units, $\epsilon_0=\mu_0=1$. A typical approach to tackle this high-dimensional problem, known as topology optimization (TopOpt), is to consider a material interpolation/relaxation procedure allowing for fast calculations of objective gradients through adjoints along with filtering techniques that binarize the resulting designs~\cite{christiansen_inverse_2021, schubert_strict_fabrication_2022}. 
Since $\chi(\bm r)$ enters the optimization program as a nonlinear constraint, the problem is nonconvex and direct certification of optimal solutions is not possible beyond extremely simple cases.

A general method for bounding \eqref{eq:topopt} is to convert the optimization problem from material to polarization-field degrees of freedom $\bm P = \chi(\bm r) \bm E_t$; such a transformation places no restrictions on the structure given that for a fixed $\bm J_i$, any $\chi(\bm r)$ will have a corresponding $\bm P$. 
Let $\G_0$ be the Green's function operator for Maxwell's equations in vacuum, $\left(\nabla \times \nabla \times - \omega^2 \right) \G_0 = \omega^2 \mI$ with $\mI$ the identity operator. The total electric field may be decomposed into its incident and scattered components $\bm E_t = \bm E_i +\bm E_s$, with the incident vacuum field $\bm E_i = (i/\omega) \G_0 \bm J_i$ and the scattered field $\bm E_s = \G_0 \bm P$. 
The design objective $f(\cdot)$ can thus be equally well expressed using $\bm P$; crucially, $\bm P$ implicitly contains information on the device geometry and satisfies certain shape-agnostic quadratic constraints, resulting in the field optimization problem
\begin{align*}
    \max_{\bm P} \quad &f(\bm P) 
    \numthis\label{eq:QCQP} \\
    \textrm{s.t.} \quad &\forall \Omega_j \subseteq \Omega, 
    \\
    & \underbrace{\int_{\Omega_j} d \bm r \bm P^* (\bm r) \left[ \left( \left( \chi^{-1} - \G \right) \bm P \right) (\bm r) - \bm E_i(\bm r) \right]}_{ \equiv C_j(\bm P)} = 0. 
\end{align*} 
The constraints $C_j$ represent power conservation in any sub-region $\Omega_j$ and depend solely on the fixed susceptibility $\chi$ and vacuum Green's function $\G_0$. They can be derived using Poynting's theorem~\cite{chaoMaximumElectromagnetic2022}, or alternatively by taking inner products over the binary choice that at any given $\bm r$, either $\bm P=0$ or $\chi^{-1} \bm P = \bm E_i + \G_0 \bm P$~\cite{gertlerManyPhotonic2025}. If all possible $\Omega_j \subseteq \Omega$ are used for constraints, then \eqref{eq:QCQP} and \eqref{eq:topopt} are equivalent~\cite{molesky_t_communication_2022,gertlerManyPhotonic2025}: in practice this corresponds to having a constraint for every pixel of a discretization of $\Omega$. Including fewer constraints results in a simpler problem that gives an upper bound to \eqref{eq:topopt}. 

While \eqref{eq:QCQP} is still non-convex, it is a QCQP if $f(\bm P)$ is a quadratic function and can be bounded with well-established convex relaxation techniques. Namely, defining the Lagrangian as $\Lag(\bm \phi,\bm P) \equiv f(\bm P) + \sum \phi_j C_j(\bm P)$, the dual function over the Lagrange multipliers $\dual(\bm \phi) = \max_{\bm P} \Lag(\bm \phi, \bm P)$ is a bound on \eqref{eq:QCQP}~\cite{boyd_convex_2004}. Weak duality refers to the fact that $\dual(\bm \phi)$ always gives an upper bound to \eqref{eq:QCQP}, and hence to \eqref{eq:topopt}. We find the tightest such bound by solving the dual problem
\begin{equation}
    \min_{\bm \phi} \quad \dual(\bm \phi) = \min_{\bm \phi} \max_{\bm P} \quad f(\bm P) + \sum \phi_j C_j(\bm P)\label{eq:dualopt}
\end{equation}
which is guaranteed to be convex~\cite{boyd_convex_2004}; when \eqref{eq:QCQP} is a QCQP, $\dual$ and its derivatives have simple analytical expressions~\cite{molesky_global_2020,chaoProbingFundamental2023,amaolo_can_heterostructures_2023}, allowing for the use of gradient and Hessian based optimization methods. Further numerical details on solving the dual problem efficiently are described in the Appendix. 

Denote the multiplier solution to the dual problem as $\bm \phi_{\dual} \equiv \arg\min_\phi \dual(\phi)$ and the corresponding dual optimal polarization as $\bm P_\dual \equiv \arg \max_{\bm P} \Lag(\phi_\dual, \bm P)$. If $\bm P_\dual$ satisfies all constraints $C_j$, then it is the global optimum of \eqref{eq:QCQP}, a condition known as strong duality. Strong duality generally does not hold when \eqref{eq:QCQP} is non-convex; however, since $\bm P_{\dual}$ originates from the dual convex relaxation, one might expect that it nevertheless contains information on desirable field characteristics for optimal structures. To infer a material structure from a given dual polarization distribution, we exploit the defining relation between polarization and field response:
\begin{equation} \label{eq:inference}
    \chi_\textrm{inf}(\bm r) = \frac{\bm P_{\dual} (\bm r)}{ \bm E_t (\bm r)} = \dfrac{\bm P_{\dual} (\bm r)}{\bm E_i + (\G_0\bm P_{\dual})(\bm r)}.
\end{equation}
Since $\bm P_{\dual}$ can violate a subset of the power conservation constraints, $\chi_\textrm{inf}(\bm r)$ does not conform to the material specifications of \eqref{eq:topopt}: we thus project $\chi_\textrm{inf}(\bm r)$ onto a suitable material interpolation as a design template for further refinement using TopOpt. For this work, we use a simple linear interpolation $\chi(\bm r) = \rho(\bm r) \chi$ with $\rho(\bm r) \in [0,1]$, and consider an initialization $\rho_0(\bm r)$ for TopOpt given by clipped projection:
\begin{equation}
 \rho_0(\bm r) = \begin{cases}
    0, & \frac{\Re{\chi_\textrm{inf}(\bm r)^{*}\chi}}{\left|\chi\right|^{2}} < 0 \\
    \frac{\Re{\chi_\textrm{inf}(\bm r)^{*}\chi}}{\left|\chi\right|^{2}}, & 0 \leq \frac{\Re{\chi_\textrm{inf}(\bm r)^{*}\chi}}{\left|\chi\right|^{2}} < 1 \\
    1, & 1 \leq \frac{\Re{\chi_\textrm{inf}(\bm r)^{*}\chi}}{\left|\chi\right|^{2}}
\end{cases}. 
\label{eq:inference}
\end{equation}

\begin{figure*}[ht]
    \centering
    \includegraphics[width=\linewidth]{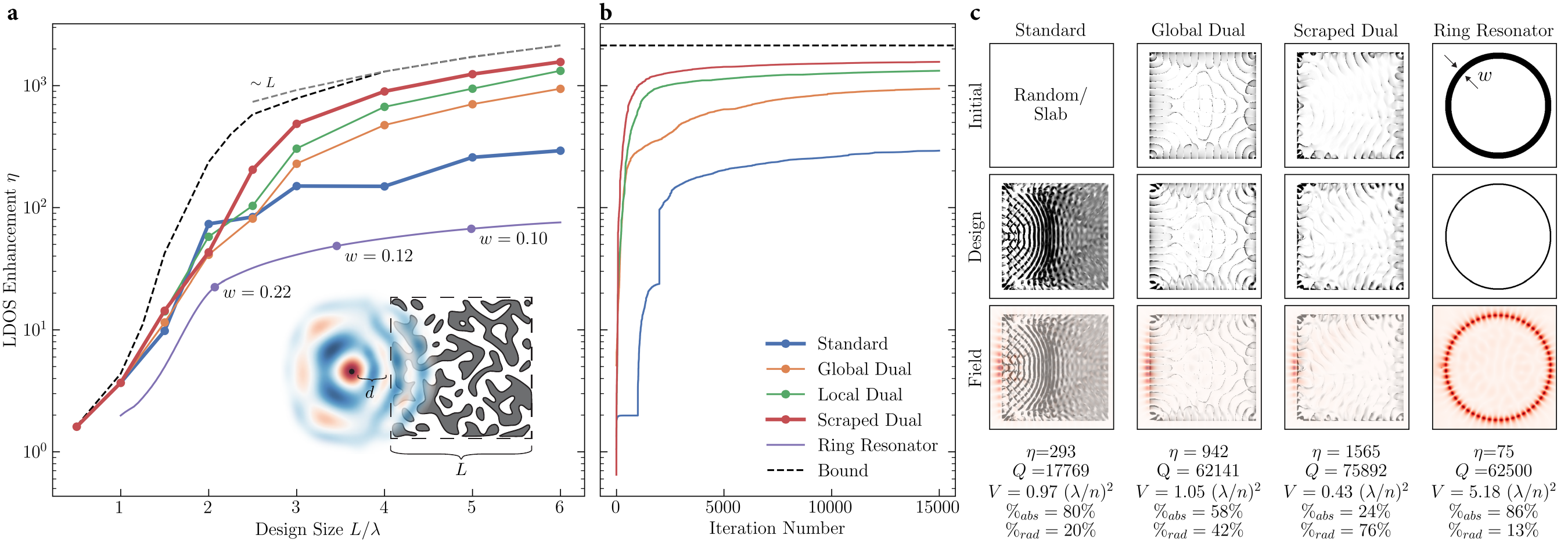}
    \caption{\textbf{Approaching the limits of Purcell enhancement via verlan design}. (a) Scaling of LDOS enhancement as a function of side-length for a point source a distance \(d=0.2\lambda\) from a \(L\times L\) design region with design material \(\chi = 5+10^{-4}i\). Dashed lines show local dual bounds; solid lines show the performance of various inverse designed structures. 
    All ring resonators have optimized width \(w\) for a given $L$ (although only a few values of \(w\) are noted). 
    Standard inverse design is done using TopOpt with ten random, vacuum, 1/2-slab, and full-slab initializations. 
    The verlan designs originate from global dual bounds (orange), local dual bounds (green), and scraping the local dual to approach strong duality (red)~\cite{molesky_verlan}. 
    Standard TopOpts for $L\geq 4$ achieved its best results using $Q$-ramping (see main text) with the Lorentzian lineshape $Q=\{10, 10^3, \infty\}$ progressively ramped up every $10^3$ iterations; TopOpt after verlan extraction is always performed at the real frequency $\omega$. 
    (b) LDOS enhancement as a function of TopOpt iteration number for the \(L=6\lambda\) designs.
    Results in (a) and (b) show the best verlan designs converging to performance within a factor of 2 of optimal and about 5 times that of standard TopOpt, in significantly fewer iterations.
    (c) TopOpt initializations, final designs, and dipole fields for all the cases described above at \(L=6\lambda\), with performance data for the final designs: LDOS enhancement relative to vacuum $\eta$, mode quality factor \(Q\), mode volume \(V\), as well as the percent of power extracted attributed to absorption and radiation. 
    Verlan structures have higher $Q$, smaller $V$, and fundamentally different enhancement mechanisms (see Fig.~\ref{fig:mechanism}) than standard TopOpt.}
    \label{fig:TOresults}
\end{figure*}
Although other inference relations could be sensibly considered, this simple form is exact in the case of strong duality, minimizes Euclidean distance when $\chi_\textrm{inf}(\bm r)$ does not conform to material specifications, and is capable of guiding topology optimization towards near-optimal designs (as demonstrated by the results shown below). 
If $\chi_\textrm{inf}(\bm r)$ deviates strongly from imposed material specifications, we believe that instead of proposing more complicated inference rules, it is more productive to modify the underlying QCQP towards strongly duality: the best inference rule is a function of the design problem. 
Our procedure for carrying out these modifications, detailed in Ref.~\cite{molesky_verlan}, is to refine the linear term of the design objective $f(\bm P)$ through a process we refer to as ``scraping.''
Under the application of a scrape, the dual moves towards strong duality, leading $\chi_\textrm{inf}(\bm r)$ to automatically align with material specifications. 


\section{Results}
Fig.~\ref{fig:TOresults} reports the application of our verlan design framework to the problem of maximizing the Purcell enhancement of a 2D single-frequency emitter in the vicinity of a structured photonic device of material susceptibility $\chi$ contained within a square design region of size $L\times L$. 
By Poynting's theorem, one can relate Purcell enhancement to the photonic local density of states (LDOS), which is proportional to the total extinguished power of a point source~\cite{novotnyPrinciplesNanooptics2012},  $\rho(\bm E_t) = -\frac{1}{2}\Re{\bm J_i \cdot \bm E_t}$ with $\bm J_i(\bm r) = \delta(\bm r - \bm r_{source})$~\cite{shim_fundamental_2019,chaoMaximumElectromagnetic2022}. Intrinsically, the LDOS is influenced by the presence and accessibility of photonic modes, i.e., source-free solutions to Maxwell's equations,
\begin{equation}
    \nabla \times \nabla \times \bm E_m - (\bm 1+\chi(\bm r)) \omega_m^2 \bm E_m = 0
\end{equation}
where the resonance frequency $\omega_m$ is generally complex due to absorptive and radiative losses. Large LDOS can be achieved in structures supporting temporally long-lived and spatially localized modes with high quality factors $Q = \frac{\Re{\omega_m}}{2\Im{\omega_m}}$ and small mode volumes $V = \frac{\int d^2\bm r \, \Re{\epsilon(\bm r)} |\bm{E}_m(\bm r)|^2}{\max_{\bm r} \Re{\epsilon(\bm r)} |\bm{E}_m(\bm r)|^2}$ that are on resonance at the target frequency ($|\omega_m - \omega| < \omega/Q $).
These properties lead to an increased sensitivity to structural changes: achieving high-$Q$ resonances in a compact domain requires careful interference engineering to cancel radiative losses and minimize absorption~\cite{johnsonMultipolecancellationMechanism2001,srinivasanMomentumSpace2002,seokRadiationEngineering2011,wangHighQPlasmonic2021}; decreasing linewidths means that small structural perturbations can easily shift the mode off resonance~\cite{katoPerturbationTheory1995,johnsonPerturbationTheory2002,loganSelectiveActive2024}. 
Consequently, the structural optimization problem is ill-conditioned, with local optima resting on narrow ``knife-edge’’ ridges in the design space that can both trap and slow gradient-based methods~\cite{liangFormulationScalable2013}. 

A standard technique to mitigate such ill-conditioning is to optimize for LDOS averaged over a Lorentzian spectral lineshape~\cite{liangFormulationScalable2013,wang_maximizing_2018}: beginning with a broadband window, the linewidth is gradually reduced, effectively throttling the operating $Q$ and biasing the optimization towards increasing field localization at the emitter. While this $Q$-ramping procedure has been successfully applied to the design of ultra-small $V$ cavities~\cite{albrechtsenNanometerscalePhoton2022}, for the external emitter configuration considered in this paper there is a non-trivial trade-off between large field localization at the target point (by concentrating field outside the structure) and higher $Q$ (by concentrating field within the design bulk). 
When $Q$-ramping, the initial broad bandwidth optimization may over-commit to decreasing $V$ at the expense of $Q$: we have observed in recent work that for large design sizes and low-loss materials, simple ring resonators supporting high-$Q$, delocalized whispering gallery modes can outperform $Q$-ramped TopOpt~\cite{chaoMaximumElectromagnetic2022}. 

Given these challenges, the maximization of Purcell enhancement is well suited for benchmarking our verlan approach against both standard TopOpt and intuitive ring-resonator designs. Figure.~\ref{fig:TOresults} shows LDOS enhancements obtained structuring a low-index, lossy dielectric $\chi = 5 + 10^{-4}i$ over varying design sizes $L$. TopOpt results were obtained by selecting the best of several structures obtained with and without $Q$-ramping, starting with homogeneous initializations $\rho_0 = \bm 1$, $\rho_0 = \bm 0$, $\rho_0 = 0.5 \cdot \bm 1$ as well as ten random $\rho_0 \sim \text{Unif}[0,1]^N$. For ring resonators, the inner and outer radii were optimized for best performance. For verlan design, we extracted TopOpt $\rho_0$ from dual bounds incorporating various degrees of physical relaxations enforced by the volumetric power-conservation constraints, including either a single constraint over the entire design region $\Omega$ (global dual), or all possible pixel-level constraints (local dual). Also shown are verlan results from bounds with pixel-level constraints on objectives modified to approach strong duality (scraped dual), see Ref.~\cite{molesky_verlan}.
All Maxwell calculations for TopOpts and bounds were performed using an in-house FDFD solver on a square grid with resolution $\lambda/40$. TopOpts were performed using the method of moving asymptotes algorithm~\cite{svanbergMethodMoving1987a} as implemented in the open source package NLopt~\cite{johnson_nlopt_2019}, run until appropriately converged or to a fixed $15000$ iterations if convergence is slow due to ill-conditioning. Note the lack of feature size filtering or binarization of TopOpt designs, which will naturally lead to some degradation of performance; alternatively, intermediate values of $\rho(\bm r)$ may be interpreted as representing sub-pixel sized structural features through material homogenization~\cite{smithHomogenizationMetamaterials2006,tsukermanEffectiveParameters2011,markel_maxwell_2016}. 

As seen in Figure.~\ref{fig:TOresults}, verlan designs dramatically outperform ring resonators as well as standard TopOpt geometries, particularly at larger system sizes ($L\geq 2.5\lambda$): the scraped dual designs achieve an improvement factor $>5$ compared to standard TopOpt, and approach the fundamental performance limits to within a factor of two. Remarkably, while global dual bounds incorporate only a single constraint---and are therefore significantly faster to compute---they are seen to also outperform standard TopOpt by around a factor of three. This trend both showcases the relevance of global wave considerations for optimal design and suggests immense potential for scaling verlan design to larger domain sizes. The superiority of the verlan designs can also be seen in the optimization progressions shown in Figure~\ref{fig:TOresults}b, with verlan initializations rapidly converging to better performing solutions without any need for $Q$-ramping. Notably, verlan structures also exhibit qualitatively different geometric features: standard TO designs have high material fill fractions, with the structure close to the dipole resembling  half of a bullseye grating~\cite{butcherAlldielectricMultiresonant2022}. Verlan designs, in contrast, exhibit low fill fractions and a strong distinction between the geometric patterns in the bulk and surface. 

\begin{figure}
    \centering
    \includegraphics[width=\linewidth]{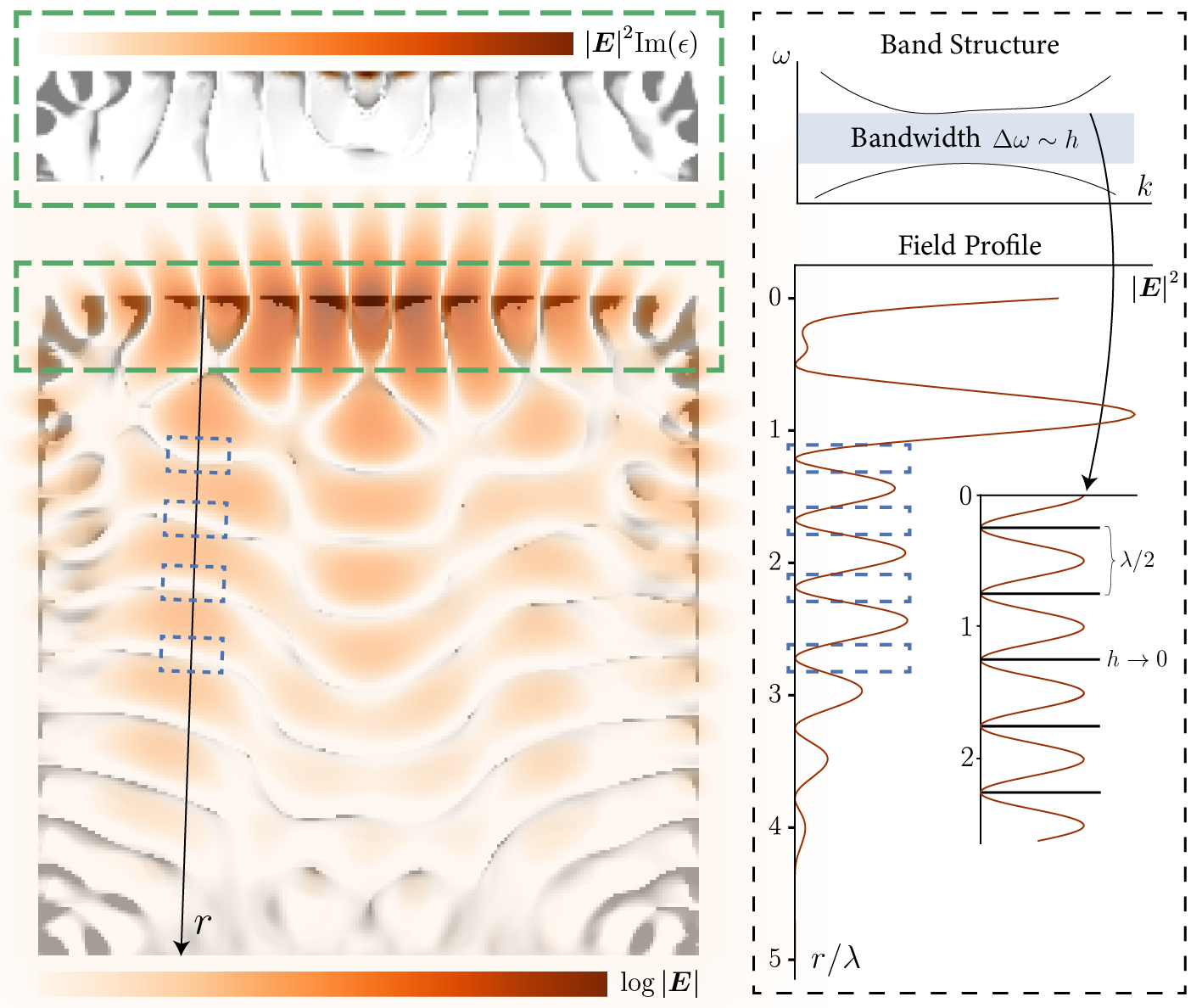}
    \caption{\textbf{Physical mechanisms behind near-optimal Purcell enhancement in verlan designs}. The structure and field profiles shown correspond to scraped dual verlan designs for $L=5\lambda$, and exhibit distinct features between the surface and bulk. At the structure surface, there is a thin waveguide-like strip with vertical grating patterns extending into the bulk (green boxes) that concentrates the field using pseudogap confinement: a combination of index guiding in the vertical direction and Bragg reflection in the horizontal direction to center the mode close to the surface. The absorbed power distribution (top green box) shows that most absorption is concentrated in the surface around the target point. The $\log(|\bm{E}|)$ distribution (lower left) further corroborates this, showing the field corralled in vacuum pockets by strips of thin gratings running across the bulk of the structure (blue boxes). These gratings have period around $\lambda/2$ with small grating thickness $h$ and confine the radiation leakage from the surface pseudogap in a delocalized, loss-minimizing fashion; see main text for details. 
    }
    \label{fig:mechanism}
\end{figure}

Inspection of the modal field profiles indicates that the balancing of delocalization and loss is the central theme underlying the varying geometric features and performance gaps between different designs. While the ring resonators have exponentially decaying radiative losses as $L$ increases~\cite{marcatili_bends_1969,joannopoulos_photonic_2008}, the whispering gallery modes they support are highly delocalized (large $V$) and also overlap significantly with the lossy material, leading to absorption that restricts further growth in $Q$. The optimal ring width is thinner for larger $L$ in order to spread the mode radially and reduce absorption. On the other hand, standard TopOpt wih $Q$-ramping creates tightly-localized resonances, but the $Q$ suffers: specifically for $L=6$ as shown in Figure.~\ref{fig:TOresults}, standard TopOpt has subwavelength $V=0.97(\lambda/n)^2$ which is less than one fifth of the ring resonator $V=5.18 (\lambda/n)^2$, while ring resonator $Q=6.25\times 10^4$ is about $3.5$ times that of the standard TopOpt $Q=1.78\times10^4$.  The verlan designs enjoy the best of both worlds, having long-lived, highly concentrated resonances with $Q$ comparable to that of ring resonators and $V$ comparable to that of standard TopOpt. This results in a faster performance scaling with $L$ that keeps pace with the growth of the bounds, unlike the other two approaches. 

Figure~\ref{fig:mechanism} zooms in on the $L=5\lambda$ scraped dual verlan design and its dominant mode, depicting its LDOS enhancement mechanism in detail. The top layer of the structure consists of a narrow waveguide with a vertical grating: this is akin to the pseudogap confinement of nanobeam cavities~\cite{quanDeterministicDesign2011,seidlerSlottedPhotonic2013} with index guiding in the vertical direction and Bragg reflection in the horizontal direction. While pseudogap confinement is effective at anchoring the mode near the target source, its main bottleneck is radiation leaking in the directions without a bandgap. The vertical grating preferentially channels this radiation into the structure bulk, where it is further contained with minimal absorption by a collection of thin curved gratings. The working principle of these curved gratings is based on air band slow light, as first discussed in Ref.~\cite{strekhaSuppressingElectromagnetic2024}: the period of a 1D infinite grating is set at $\lambda/2$ while the material layer thickness $h$ is reduced to approach 0, situating $\omega$ at the bottom of the air band where there is a standing wave. The mode field zeros are located within the material, and $|\bm E|^2 \propto r^2$ a small distance $r$ away from a zero, so the absorption per unit cell vanishes $\propto \int_0^h r^2 dr \propto h^3$. Combined with a diverging impedance mismatch between free space propagation and the slow light standing wave, a half-infinite 1D grating is able to achieve perfect reflectivity despite the absorptive material~\cite{strekha_suppressing_2024}. The field profile in the verlan structure bulk echoes this idealized 1D analysis, with compromises to account for the 2D compact design domain such as grating curvature, a finite $h$ and adiabatic thickening of the material near the boundary. Overall, absorption is concentrated within the surface waveguide, while the fields over the remainder of the structure are delicately corralled within vacuum pockets between thin material strands forming an interlocking Bragg network.

\section{Discussion}

We have presented a novel verlan design framework for photonics that exploits optimal field information obtained from fundamental limits to inform inverse design. Applied to the challenging problem of maximizing Purcell enhancement, verlan design discovered wavelength scale structures that operate near the upper bounds of achievable performance. Through a harmonious combination of index guiding~\cite{yariv_photonics_2006}, pseudogap confinement~\cite{quanDeterministicDesign2011}, and air band slow light (a special case of 1D Bragg scattering favoring field power in vacuum)~\cite{strekha_suppressing_2024}, the verlan structures support high-$Q$ resonances with strong field concentration near the surface, vastly outperforming intuitive ring resonators and standard inverse design. This demonstrates the exceptional capability of verlan design in balancing field localization and absorption, a topic especially relevant for applications involving metals and lossy dielectrics~\cite{boriskinaLossesPlasmonics2017}. 

Moving beyond the 2D illustrative example discussed here, application of verlan design to even larger practical settings---involving design regions $\gtrsim 100\lambda^2$ in 2D and $\gtrsim \lambda^3$ in 3D---demands further improvements to computational efficiency. Given the highly unfavorable time and memory scaling of general interior-point SDP solvers~\cite{ahmadiDSOSSDSOS2019} as well as the ``no free lunch'' maxim of adapting algorithms to particular problems~\cite{wolpertNoFree1997,hoSimpleExplanation2002}, we believe the key to further computational improvements lies in exploiting the special mathematical structure of photonic design QCQPs. 
Specifically, the common appearance of products of vacuum Green's functions with spatial projections can lead to convex optimization speedups such as the generalized constraint descent we employed in this paper (Appendix~\ref{asec:gcd}). 
Algebraic properties of the vacuum Green's function such as reciprocity and translation invariance already enable large-scale Maxwell solvers through techniques such as preconditioning~\cite{aruliahMultigridPreconditioning2002,vionImprovedSweeping2018,bonazzoliDomainDecomposition2019} and the Fast Fourier Transform~\cite{zhangFFTbasedAlgorithm2018,harrisFastLargeScale2019}; it should be relatively straightforward to incorporate these into verlan design. 
Additional efficiency gains could be achieved by reducing the number of constraints enforced during the structure extraction process: as seen in the present examples, using only global power conservation for verlan design also lead to large performance gains. This echoes earlier results concerning limits on absorption cross sections~\cite{molesky_global_2020}, power extraction~\cite{chaoMaximumElectromagnetic2022}, and Raman scattering~\cite{chaoSumofSquaresBounds2025}, where the imposition of global constraints alone was sufficient to produce nearly tight bounds for sufficiently large structures. Further study is needed to understand to what extent this observation remains true across a wider variety of photonic problems. 

Given the promising results demonstrated in this paper and other related works, we anticipate that application of the verlan approach to other challenging photonics design problems~\cite{chenValidationCharacterization2024} will yield further novel insights. Verlan design may also be useful for inverse design in other physical domains where the QCQP convex relaxation framework is applicable such as quantum control~\cite{zhangConservationlawbasedGlobal2021} and elasticity~\cite{dalklintPerformanceBounds2024}. Multiphysics analysis is also possible, and may be especially relevant for imposing fabrication-related restrictions such as minimum feature size~\cite{lazarovFiltersTopology2011} and connectivity constraints~\cite{kusterInverseDesign2025} through the use of auxiliary field variables. Finally, given that other inverse design methods may still out-perform verlan design for certain examples and initializations, it is clear that further improvements to verlan structure extraction is possible. An intriguing direction is to consider whether existing designs can be incorporated into the verlan extraction process to provide further information on the optimization landscape, moving towards a new paradigm where inverse design and fundamental limits calculations seamlessly integrate in pursuit of certified optimal photonics engineering.

\bibliography{refs,verlan_short_refs} 

\begin{thebibliography}{85}%
\makeatletter
\providecommand \@ifxundefined [1]{%
 \@ifx{#1\undefined}
}%
\providecommand \@ifnum [1]{%
 \ifnum #1\expandafter \@firstoftwo
 \else \expandafter \@secondoftwo
 \fi
}%
\providecommand \@ifx [1]{%
 \ifx #1\expandafter \@firstoftwo
 \else \expandafter \@secondoftwo
 \fi
}%
\providecommand \natexlab [1]{#1}%
\providecommand \enquote  [1]{``#1''}%
\providecommand \bibnamefont  [1]{#1}%
\providecommand \bibfnamefont [1]{#1}%
\providecommand \citenamefont [1]{#1}%
\providecommand \href@noop [0]{\@secondoftwo}%
\providecommand \href [0]{\begingroup \@sanitize@url \@href}%
\providecommand \@href[1]{\@@startlink{#1}\@@href}%
\providecommand \@@href[1]{\endgroup#1\@@endlink}%
\providecommand \@sanitize@url [0]{\catcode `\\12\catcode `\$12\catcode `\&12\catcode `\#12\catcode `\^12\catcode `\_12\catcode `\%12\relax}%
\providecommand \@@startlink[1]{}%
\providecommand \@@endlink[0]{}%
\providecommand \url  [0]{\begingroup\@sanitize@url \@url }%
\providecommand \@url [1]{\endgroup\@href {#1}{\urlprefix }}%
\providecommand \urlprefix  [0]{URL }%
\providecommand \Eprint [0]{\href }%
\providecommand \doibase [0]{https://doi.org/}%
\providecommand \selectlanguage [0]{\@gobble}%
\providecommand \bibinfo  [0]{\@secondoftwo}%
\providecommand \bibfield  [0]{\@secondoftwo}%
\providecommand \translation [1]{[#1]}%
\providecommand \BibitemOpen [0]{}%
\providecommand \bibitemStop [0]{}%
\providecommand \bibitemNoStop [0]{.\EOS\space}%
\providecommand \EOS [0]{\spacefactor3000\relax}%
\providecommand \BibitemShut  [1]{\csname bibitem#1\endcsname}%
\let\auto@bib@innerbib\@empty
\bibitem [{\citenamefont {Yariv}\ and\ \citenamefont {Yeh}(2006)}]{yariv_photonics_2006}%
  \BibitemOpen
  \bibfield  {author} {\bibinfo {author} {\bibfnamefont {A.}~\bibnamefont {Yariv}}\ and\ \bibinfo {author} {\bibfnamefont {P.}~\bibnamefont {Yeh}},\ }\href@noop {} {\emph {\bibinfo {title} {Photonics: Optical Electronics in Modern Communications}}}\ (\bibinfo  {publisher} {{Oxford University Press, Inc.}},\ \bibinfo {year} {2006})\BibitemShut {NoStop}%
\bibitem [{\citenamefont {Joannopoulos}\ \emph {et~al.}(2008)\citenamefont {Joannopoulos}, \citenamefont {Steven}, \citenamefont {Winn},\ and\ \citenamefont {Meade}}]{joannopoulos_photonic_2008}%
  \BibitemOpen
  \bibfield  {author} {\bibinfo {author} {\bibfnamefont {J.~D.}\ \bibnamefont {Joannopoulos}}, \bibinfo {author} {\bibfnamefont {J.~G.}\ \bibnamefont {Steven}}, \bibinfo {author} {\bibfnamefont {J.~N.}\ \bibnamefont {Winn}},\ and\ \bibinfo {author} {\bibfnamefont {R.~D.}\ \bibnamefont {Meade}},\ }\href@noop {} {\emph {\bibinfo {title} {Photonic Crystals: Molding the Flow of Light}}},\ \bibinfo {edition} {2nd}\ ed.\ (\bibinfo  {publisher} {{Princeton University Press}},\ \bibinfo {address} {{Princeton}},\ \bibinfo {year} {2008})\BibitemShut {NoStop}%
\bibitem [{\citenamefont {Maier}(2007)}]{maier_plasmonics_2007}%
  \BibitemOpen
  \bibfield  {author} {\bibinfo {author} {\bibfnamefont {S.~A.}\ \bibnamefont {Maier}},\ }\href@noop {} {\emph {\bibinfo {title} {Plasmonics: Fundamentals and Applications}}}\ (\bibinfo  {publisher} {{Springer Science \& Business Media}},\ \bibinfo {year} {2007})\BibitemShut {NoStop}%
\bibitem [{\citenamefont {Roberts}\ \emph {et~al.}(2023)\citenamefont {Roberts}, \citenamefont {Ballew}, \citenamefont {Zheng}, \citenamefont {Garcia}, \citenamefont {Camayd-Muñoz}, \citenamefont {Hon},\ and\ \citenamefont {Faraon}}]{roberts_3d_2023}%
  \BibitemOpen
  \bibfield  {author} {\bibinfo {author} {\bibfnamefont {G.}~\bibnamefont {Roberts}}, \bibinfo {author} {\bibfnamefont {C.}~\bibnamefont {Ballew}}, \bibinfo {author} {\bibfnamefont {T.}~\bibnamefont {Zheng}}, \bibinfo {author} {\bibfnamefont {J.~C.}\ \bibnamefont {Garcia}}, \bibinfo {author} {\bibfnamefont {S.}~\bibnamefont {Camayd-Muñoz}}, \bibinfo {author} {\bibfnamefont {P.~W.~C.}\ \bibnamefont {Hon}},\ and\ \bibinfo {author} {\bibfnamefont {A.}~\bibnamefont {Faraon}},\ }\bibfield  {title} {\bibinfo {title} {3d-patterned inverse-designed mid-infrared metaoptics},\ }\href {https://doi.org/10.1038/s41467-023-38258-2} {\bibfield  {journal} {\bibinfo  {journal} {Nature Communications}\ }\textbf {\bibinfo {volume} {14}},\ \bibinfo {pages} {2768} (\bibinfo {year} {2023})}\BibitemShut {NoStop}%
\bibitem [{\citenamefont {Jensen}\ and\ \citenamefont {Sigmund}(2011)}]{jensenTopologyOptimization2011}%
  \BibitemOpen
  \bibfield  {author} {\bibinfo {author} {\bibfnamefont {J.~S.}\ \bibnamefont {Jensen}}\ and\ \bibinfo {author} {\bibfnamefont {O.}~\bibnamefont {Sigmund}},\ }\bibfield  {title} {\bibinfo {title} {Topology optimization for nano-photonics},\ }\href@noop {} {\bibfield  {journal} {\bibinfo  {journal} {Laser \& Photonics Reviews}\ }\textbf {\bibinfo {volume} {5}},\ \bibinfo {pages} {308} (\bibinfo {year} {2011})}\BibitemShut {NoStop}%
\bibitem [{\citenamefont {Molesky}\ \emph {et~al.}(2018)\citenamefont {Molesky}, \citenamefont {Lin}, \citenamefont {Piggott}, \citenamefont {Jin}, \citenamefont {Vu{\v c}kovi{\'c}},\ and\ \citenamefont {Rodriguez}}]{molesky_inverse_2018}%
  \BibitemOpen
  \bibfield  {author} {\bibinfo {author} {\bibfnamefont {S.}~\bibnamefont {Molesky}}, \bibinfo {author} {\bibfnamefont {Z.}~\bibnamefont {Lin}}, \bibinfo {author} {\bibfnamefont {A.~Y.}\ \bibnamefont {Piggott}}, \bibinfo {author} {\bibfnamefont {W.}~\bibnamefont {Jin}}, \bibinfo {author} {\bibfnamefont {J.}~\bibnamefont {Vu{\v c}kovi{\'c}}},\ and\ \bibinfo {author} {\bibfnamefont {A.~W.}\ \bibnamefont {Rodriguez}},\ }\bibfield  {title} {\bibinfo {title} {Inverse design in nanophotonics},\ }\href@noop {} {\bibfield  {journal} {\bibinfo  {journal} {Nature Photonics}\ }\textbf {\bibinfo {volume} {12}},\ \bibinfo {pages} {659} (\bibinfo {year} {2018})}\BibitemShut {NoStop}%
\bibitem [{\citenamefont {Li}\ \emph {et~al.}(2022)\citenamefont {Li}, \citenamefont {Pestourie}, \citenamefont {Lin}, \citenamefont {Johnson},\ and\ \citenamefont {Capasso}}]{liEmpoweringMetasurfaces2022}%
  \BibitemOpen
  \bibfield  {author} {\bibinfo {author} {\bibfnamefont {Z.}~\bibnamefont {Li}}, \bibinfo {author} {\bibfnamefont {R.}~\bibnamefont {Pestourie}}, \bibinfo {author} {\bibfnamefont {Z.}~\bibnamefont {Lin}}, \bibinfo {author} {\bibfnamefont {S.~G.}\ \bibnamefont {Johnson}},\ and\ \bibinfo {author} {\bibfnamefont {F.}~\bibnamefont {Capasso}},\ }\bibfield  {title} {\bibinfo {title} {Empowering {{Metasurfaces}} with {{Inverse Design}}: {{Principles}} and {{Applications}}},\ }\href {https://doi.org/10.1021/acsphotonics.1c01850} {\bibfield  {journal} {\bibinfo  {journal} {ACS Photonics}\ }\textbf {\bibinfo {volume} {9}},\ \bibinfo {pages} {2178} (\bibinfo {year} {2022})}\BibitemShut {NoStop}%
\bibitem [{\citenamefont {{Lalau-Keraly}}\ \emph {et~al.}(2013)\citenamefont {{Lalau-Keraly}}, \citenamefont {Bhargava}, \citenamefont {Miller},\ and\ \citenamefont {Yablonovitch}}]{lalau-keraly_adjoint_2013}%
  \BibitemOpen
  \bibfield  {author} {\bibinfo {author} {\bibfnamefont {C.~M.}\ \bibnamefont {{Lalau-Keraly}}}, \bibinfo {author} {\bibfnamefont {S.}~\bibnamefont {Bhargava}}, \bibinfo {author} {\bibfnamefont {O.~D.}\ \bibnamefont {Miller}},\ and\ \bibinfo {author} {\bibfnamefont {E.}~\bibnamefont {Yablonovitch}},\ }\bibfield  {title} {\bibinfo {title} {Adjoint shape optimization applied to electromagnetic design},\ }\href {https://doi.org/10.1364/OE.21.021693} {\bibfield  {journal} {\bibinfo  {journal} {Optics Express}\ }\textbf {\bibinfo {volume} {21}},\ \bibinfo {pages} {21693} (\bibinfo {year} {2013})}\BibitemShut {NoStop}%
\bibitem [{\citenamefont {Hughes}\ \emph {et~al.}(2018)\citenamefont {Hughes}, \citenamefont {Minkov}, \citenamefont {Williamson},\ and\ \citenamefont {Fan}}]{hughes_adjoint_2018}%
  \BibitemOpen
  \bibfield  {author} {\bibinfo {author} {\bibfnamefont {T.~W.}\ \bibnamefont {Hughes}}, \bibinfo {author} {\bibfnamefont {M.}~\bibnamefont {Minkov}}, \bibinfo {author} {\bibfnamefont {I.~A.~D.}\ \bibnamefont {Williamson}},\ and\ \bibinfo {author} {\bibfnamefont {S.}~\bibnamefont {Fan}},\ }\bibfield  {title} {\bibinfo {title} {Adjoint {Method} and {Inverse} {Design} for {Nonlinear} {Nanophotonic} {Devices}},\ }\href {https://doi.org/10.1021/acsphotonics.8b01522} {\bibfield  {journal} {\bibinfo  {journal} {ACS Photonics}\ }\textbf {\bibinfo {volume} {5}},\ \bibinfo {pages} {4781} (\bibinfo {year} {2018})},\ \bibinfo {note} {publisher: American Chemical Society}\BibitemShut {NoStop}%
\bibitem [{\citenamefont {Piggott}\ \emph {et~al.}(2020)\citenamefont {Piggott}, \citenamefont {Ma}, \citenamefont {Su}, \citenamefont {Ahn}, \citenamefont {Sapra}, \citenamefont {Vercruysse}, \citenamefont {Netherton}, \citenamefont {Khope}, \citenamefont {Bowers},\ and\ \citenamefont {Vu{\v c}kovi{\'c}}}]{piggottInverseDesignedPhotonics2020}%
  \BibitemOpen
  \bibfield  {author} {\bibinfo {author} {\bibfnamefont {A.~Y.}\ \bibnamefont {Piggott}}, \bibinfo {author} {\bibfnamefont {E.~Y.}\ \bibnamefont {Ma}}, \bibinfo {author} {\bibfnamefont {L.}~\bibnamefont {Su}}, \bibinfo {author} {\bibfnamefont {G.~H.}\ \bibnamefont {Ahn}}, \bibinfo {author} {\bibfnamefont {N.~V.}\ \bibnamefont {Sapra}}, \bibinfo {author} {\bibfnamefont {D.}~\bibnamefont {Vercruysse}}, \bibinfo {author} {\bibfnamefont {A.~M.}\ \bibnamefont {Netherton}}, \bibinfo {author} {\bibfnamefont {A.~S.~P.}\ \bibnamefont {Khope}}, \bibinfo {author} {\bibfnamefont {J.~E.}\ \bibnamefont {Bowers}},\ and\ \bibinfo {author} {\bibfnamefont {J.}~\bibnamefont {Vu{\v c}kovi{\'c}}},\ }\bibfield  {title} {\bibinfo {title} {Inverse-{{Designed Photonics}} for {{Semiconductor Foundries}}},\ }\href {https://doi.org/10.1021/acsphotonics.9b01540} {\bibfield  {journal} {\bibinfo  {journal} {ACS Photonics}\ }\textbf {\bibinfo {volume} {7}},\ \bibinfo {pages} {569} (\bibinfo {year} {2020})}\BibitemShut {NoStop}%
\bibitem [{\citenamefont {Hammond}\ \emph {et~al.}(2021)\citenamefont {Hammond}, \citenamefont {Oskooi}, \citenamefont {Johnson},\ and\ \citenamefont {Ralph}}]{hammondPhotonicTopology2021}%
  \BibitemOpen
  \bibfield  {author} {\bibinfo {author} {\bibfnamefont {A.~M.}\ \bibnamefont {Hammond}}, \bibinfo {author} {\bibfnamefont {A.}~\bibnamefont {Oskooi}}, \bibinfo {author} {\bibfnamefont {S.~G.}\ \bibnamefont {Johnson}},\ and\ \bibinfo {author} {\bibfnamefont {S.~E.}\ \bibnamefont {Ralph}},\ }\bibfield  {title} {\bibinfo {title} {Photonic topology optimization with semiconductor-foundry design-rule constraints},\ }\href {https://doi.org/10.1364/OE.431188} {\bibfield  {journal} {\bibinfo  {journal} {Optics Express}\ }\textbf {\bibinfo {volume} {29}},\ \bibinfo {pages} {23916} (\bibinfo {year} {2021})}\BibitemShut {NoStop}%
\bibitem [{\citenamefont {Chen}\ \emph {et~al.}(2022)\citenamefont {Chen}, \citenamefont {Lan}, \citenamefont {Su},\ and\ \citenamefont {Zhu}}]{chenInverseDesign2022}%
  \BibitemOpen
  \bibfield  {author} {\bibinfo {author} {\bibfnamefont {Y.}~\bibnamefont {Chen}}, \bibinfo {author} {\bibfnamefont {Z.}~\bibnamefont {Lan}}, \bibinfo {author} {\bibfnamefont {Z.}~\bibnamefont {Su}},\ and\ \bibinfo {author} {\bibfnamefont {J.}~\bibnamefont {Zhu}},\ }\bibfield  {title} {\bibinfo {title} {Inverse design of photonic and phononic topological insulators: A review},\ }\href {https://doi.org/10.1515/nanoph-2022-0309} {\bibfield  {journal} {\bibinfo  {journal} {Nanophotonics}\ }\textbf {\bibinfo {volume} {11}},\ \bibinfo {pages} {4347} (\bibinfo {year} {2022})}\BibitemShut {NoStop}%
\bibitem [{\citenamefont {Stich}\ \emph {et~al.}(2024)\citenamefont {Stich}, \citenamefont {Mohajan}, \citenamefont {Ceglia}, \citenamefont {Carletti}, \citenamefont {Jung}, \citenamefont {Karl}, \citenamefont {Brener}, \citenamefont {Rodriguez}, \citenamefont {Belkin},\ and\ \citenamefont {Sarma}}]{stich_inverse_2024}%
  \BibitemOpen
  \bibfield  {author} {\bibinfo {author} {\bibfnamefont {S.}~\bibnamefont {Stich}}, \bibinfo {author} {\bibfnamefont {J.}~\bibnamefont {Mohajan}}, \bibinfo {author} {\bibfnamefont {D.~d.}\ \bibnamefont {Ceglia}}, \bibinfo {author} {\bibfnamefont {L.}~\bibnamefont {Carletti}}, \bibinfo {author} {\bibfnamefont {H.}~\bibnamefont {Jung}}, \bibinfo {author} {\bibfnamefont {N.}~\bibnamefont {Karl}}, \bibinfo {author} {\bibfnamefont {I.}~\bibnamefont {Brener}}, \bibinfo {author} {\bibfnamefont {A.~W.}\ \bibnamefont {Rodriguez}}, \bibinfo {author} {\bibfnamefont {M.~A.}\ \bibnamefont {Belkin}},\ and\ \bibinfo {author} {\bibfnamefont {R.}~\bibnamefont {Sarma}},\ }\href {https://doi.org/10.48550/arXiv.2409.18196} {\bibinfo {title} {Inverse {Design} of an {All}-{Dielectric} {Nonlinear} {Polaritonic} {Metasurface}}} (\bibinfo {year} {2024}),\ \bibinfo {note} {arXiv:2409.18196 [physics]}\BibitemShut {NoStop}%
\bibitem [{\citenamefont {Roques-Carmes}\ \emph {et~al.}(2025)\citenamefont {Roques-Carmes}, \citenamefont {Wang}, \citenamefont {Yang}, \citenamefont {Majumdar},\ and\ \citenamefont {Lin}}]{roques-carmes_metaoptic_2025}%
  \BibitemOpen
  \bibfield  {author} {\bibinfo {author} {\bibfnamefont {C.}~\bibnamefont {Roques-Carmes}}, \bibinfo {author} {\bibfnamefont {K.}~\bibnamefont {Wang}}, \bibinfo {author} {\bibfnamefont {Y.}~\bibnamefont {Yang}}, \bibinfo {author} {\bibfnamefont {A.}~\bibnamefont {Majumdar}},\ and\ \bibinfo {author} {\bibfnamefont {Z.}~\bibnamefont {Lin}},\ }\bibfield  {title} {\bibinfo {title} {Metaoptic {Computational} {Imaging}},\ }\bibfield  {journal} {\bibinfo  {journal} {ACS Photonics}\ }\href {https://doi.org/10.1021/acsphotonics.4c02266} {10.1021/acsphotonics.4c02266} (\bibinfo {year} {2025}),\ \bibinfo {note} {publisher: American Chemical Society}\BibitemShut {NoStop}%
\bibitem [{\citenamefont {Angeris}\ \emph {et~al.}(2021)\citenamefont {Angeris}, \citenamefont {Vu{\v c}kovi{\'c}},\ and\ \citenamefont {Boyd}}]{angerisHeuristicMethods2021}%
  \BibitemOpen
  \bibfield  {author} {\bibinfo {author} {\bibfnamefont {G.}~\bibnamefont {Angeris}}, \bibinfo {author} {\bibfnamefont {J.}~\bibnamefont {Vu{\v c}kovi{\'c}}},\ and\ \bibinfo {author} {\bibfnamefont {S.}~\bibnamefont {Boyd}},\ }\bibfield  {title} {\bibinfo {title} {Heuristic methods and performance bounds for photonic design},\ }\href@noop {} {\bibfield  {journal} {\bibinfo  {journal} {Optics Express}\ }\textbf {\bibinfo {volume} {29}},\ \bibinfo {pages} {2827} (\bibinfo {year} {2021})}\BibitemShut {NoStop}%
\bibitem [{\citenamefont {Chao}\ \emph {et~al.}(2022{\natexlab{a}})\citenamefont {Chao}, \citenamefont {Strekha}, \citenamefont {Kuate~Defo}, \citenamefont {Molesky},\ and\ \citenamefont {Rodriguez}}]{chaoPhysicalLimits2022}%
  \BibitemOpen
  \bibfield  {author} {\bibinfo {author} {\bibfnamefont {P.}~\bibnamefont {Chao}}, \bibinfo {author} {\bibfnamefont {B.}~\bibnamefont {Strekha}}, \bibinfo {author} {\bibfnamefont {R.}~\bibnamefont {Kuate~Defo}}, \bibinfo {author} {\bibfnamefont {S.}~\bibnamefont {Molesky}},\ and\ \bibinfo {author} {\bibfnamefont {A.~W.}\ \bibnamefont {Rodriguez}},\ }\bibfield  {title} {\bibinfo {title} {Physical limits in electromagnetism},\ }\href {https://doi.org/10.1038/s42254-022-00468-w} {\bibfield  {journal} {\bibinfo  {journal} {Nature Reviews Physics}\ }\textbf {\bibinfo {volume} {4}},\ \bibinfo {pages} {543} (\bibinfo {year} {2022}{\natexlab{a}})}\BibitemShut {NoStop}%
\bibitem [{\citenamefont {Park}\ and\ \citenamefont {Boyd}(2017)}]{parkGeneralHeuristics2017}%
  \BibitemOpen
  \bibfield  {author} {\bibinfo {author} {\bibfnamefont {J.}~\bibnamefont {Park}}\ and\ \bibinfo {author} {\bibfnamefont {S.}~\bibnamefont {Boyd}},\ }\href@noop {} {\bibinfo {title} {General {{Heuristics}} for {{Nonconvex Quadratically Constrained Quadratic Programming}}}} (\bibinfo {year} {2017}),\ \Eprint {https://arxiv.org/abs/1703.07870} {arXiv:1703.07870 [math]} \BibitemShut {NoStop}%
\bibitem [{\citenamefont {Gertler}\ \emph {et~al.}(2025)\citenamefont {Gertler}, \citenamefont {Kuang}, \citenamefont {Christie}, \citenamefont {Li},\ and\ \citenamefont {Miller}}]{gertlerManyPhotonic2025}%
  \BibitemOpen
  \bibfield  {author} {\bibinfo {author} {\bibfnamefont {S.}~\bibnamefont {Gertler}}, \bibinfo {author} {\bibfnamefont {Z.}~\bibnamefont {Kuang}}, \bibinfo {author} {\bibfnamefont {C.}~\bibnamefont {Christie}}, \bibinfo {author} {\bibfnamefont {H.}~\bibnamefont {Li}},\ and\ \bibinfo {author} {\bibfnamefont {O.~D.}\ \bibnamefont {Miller}},\ }\bibfield  {title} {\bibinfo {title} {Many photonic design problems are sparse {{QCQPs}}},\ }\bibfield  {journal} {\bibinfo  {journal} {Science Advances}\ }\href {https://doi.org/10.1126/sciadv.adl3237} {10.1126/sciadv.adl3237} (\bibinfo {year} {2025})\BibitemShut {NoStop}%
\bibitem [{\citenamefont {Gustafsson}\ \emph {et~al.}(2020)\citenamefont {Gustafsson}, \citenamefont {Schab}, \citenamefont {Jelinek},\ and\ \citenamefont {Capek}}]{gustafssonUpperBounds2020}%
  \BibitemOpen
  \bibfield  {author} {\bibinfo {author} {\bibfnamefont {M.}~\bibnamefont {Gustafsson}}, \bibinfo {author} {\bibfnamefont {K.}~\bibnamefont {Schab}}, \bibinfo {author} {\bibfnamefont {L.}~\bibnamefont {Jelinek}},\ and\ \bibinfo {author} {\bibfnamefont {M.}~\bibnamefont {Capek}},\ }\bibfield  {title} {\bibinfo {title} {Upper bounds on absorption and scattering},\ }\href {https://doi.org/10.1088/1367-2630/ab83d3} {\bibfield  {journal} {\bibinfo  {journal} {New Journal of Physics}\ }\textbf {\bibinfo {volume} {22}},\ \bibinfo {pages} {073013} (\bibinfo {year} {2020})}\BibitemShut {NoStop}%
\bibitem [{\citenamefont {Molesky}\ \emph {et~al.}(2020{\natexlab{a}})\citenamefont {Molesky}, \citenamefont {Chao}, \citenamefont {Jin},\ and\ \citenamefont {Rodriguez}}]{molesky_global_2020}%
  \BibitemOpen
  \bibfield  {author} {\bibinfo {author} {\bibfnamefont {S.}~\bibnamefont {Molesky}}, \bibinfo {author} {\bibfnamefont {P.}~\bibnamefont {Chao}}, \bibinfo {author} {\bibfnamefont {W.}~\bibnamefont {Jin}},\ and\ \bibinfo {author} {\bibfnamefont {A.~W.}\ \bibnamefont {Rodriguez}},\ }\bibfield  {title} {\bibinfo {title} {Global {$\mathbb{T}$} operator bounds on electromagnetic scattering: Upper bounds on far-field cross sections},\ }\href@noop {} {\bibfield  {journal} {\bibinfo  {journal} {Physical Review Research}\ }\textbf {\bibinfo {volume} {2}},\ \bibinfo {pages} {033172} (\bibinfo {year} {2020}{\natexlab{a}})}\BibitemShut {NoStop}%
\bibitem [{\citenamefont {Kuang}\ \emph {et~al.}(2020)\citenamefont {Kuang}, \citenamefont {Zhang},\ and\ \citenamefont {Miller}}]{kuang_maximal_2020}%
  \BibitemOpen
  \bibfield  {author} {\bibinfo {author} {\bibfnamefont {Z.}~\bibnamefont {Kuang}}, \bibinfo {author} {\bibfnamefont {L.}~\bibnamefont {Zhang}},\ and\ \bibinfo {author} {\bibfnamefont {O.~D.}\ \bibnamefont {Miller}},\ }\bibfield  {title} {\bibinfo {title} {Maximal single-frequency electromagnetic response},\ }\href {https://doi.org/10.1364/OPTICA.398715} {\bibfield  {journal} {\bibinfo  {journal} {Optica}\ }\textbf {\bibinfo {volume} {7}},\ \bibinfo {pages} {1746} (\bibinfo {year} {2020})}\BibitemShut {NoStop}%
\bibitem [{\citenamefont {Molesky}\ \emph {et~al.}(2020{\natexlab{b}})\citenamefont {Molesky}, \citenamefont {Chao},\ and\ \citenamefont {Rodriguez}}]{molesky_hierarchical_2020}%
  \BibitemOpen
  \bibfield  {author} {\bibinfo {author} {\bibfnamefont {S.}~\bibnamefont {Molesky}}, \bibinfo {author} {\bibfnamefont {P.}~\bibnamefont {Chao}},\ and\ \bibinfo {author} {\bibfnamefont {A.~W.}\ \bibnamefont {Rodriguez}},\ }\bibfield  {title} {\bibinfo {title} {Hierarchical mean-field {$\mathbb{T}$} operator bounds on electromagnetic scattering: {{Upper}} bounds on near-field radiative {{Purcell}} enhancement},\ }\href {https://doi.org/10.1103/PhysRevResearch.2.043398} {\bibfield  {journal} {\bibinfo  {journal} {Physical Review Research}\ }\textbf {\bibinfo {volume} {2}},\ \bibinfo {pages} {043398} (\bibinfo {year} {2020}{\natexlab{b}})}\BibitemShut {NoStop}%
\bibitem [{\citenamefont {Kuang}\ and\ \citenamefont {Miller}(2020)}]{kuang_computational_2020}%
  \BibitemOpen
  \bibfield  {author} {\bibinfo {author} {\bibfnamefont {Z.}~\bibnamefont {Kuang}}\ and\ \bibinfo {author} {\bibfnamefont {O.~D.}\ \bibnamefont {Miller}},\ }\bibfield  {title} {\bibinfo {title} {Computational bounds to light--matter interactions via local conservation laws},\ }\href {https://doi.org/10.1103/PhysRevLett.125.263607} {\bibfield  {journal} {\bibinfo  {journal} {Physical Review Letters}\ }\textbf {\bibinfo {volume} {125}},\ \bibinfo {pages} {263607} (\bibinfo {year} {2020})}\BibitemShut {NoStop}%
\bibitem [{\citenamefont {Jelinek}\ \emph {et~al.}(2021)\citenamefont {Jelinek}, \citenamefont {Gustafsson}, \citenamefont {Capek},\ and\ \citenamefont {Schab}}]{jelinekFundamentalBounds2021}%
  \BibitemOpen
  \bibfield  {author} {\bibinfo {author} {\bibfnamefont {L.}~\bibnamefont {Jelinek}}, \bibinfo {author} {\bibfnamefont {M.}~\bibnamefont {Gustafsson}}, \bibinfo {author} {\bibfnamefont {M.}~\bibnamefont {Capek}},\ and\ \bibinfo {author} {\bibfnamefont {K.}~\bibnamefont {Schab}},\ }\bibfield  {title} {\bibinfo {title} {Fundamental bounds on the performance of monochromatic passive cloaks},\ }\href {https://doi.org/10.1364/OE.428536} {\bibfield  {journal} {\bibinfo  {journal} {Optics Express}\ }\textbf {\bibinfo {volume} {29}},\ \bibinfo {pages} {24068} (\bibinfo {year} {2021})}\BibitemShut {NoStop}%
\bibitem [{\citenamefont {Zhang}\ \emph {et~al.}(2021)\citenamefont {Zhang}, \citenamefont {Kuang}, \citenamefont {Puri},\ and\ \citenamefont {Miller}}]{zhangConservationlawbasedGlobal2021}%
  \BibitemOpen
  \bibfield  {author} {\bibinfo {author} {\bibfnamefont {H.}~\bibnamefont {Zhang}}, \bibinfo {author} {\bibfnamefont {Z.}~\bibnamefont {Kuang}}, \bibinfo {author} {\bibfnamefont {S.}~\bibnamefont {Puri}},\ and\ \bibinfo {author} {\bibfnamefont {O.~D.}\ \bibnamefont {Miller}},\ }\bibfield  {title} {\bibinfo {title} {Conservation-{{Law-Based Global Bounds}} to {{Quantum Optimal Control}}},\ }\href {https://doi.org/10.1103/PhysRevLett.127.110506} {\bibfield  {journal} {\bibinfo  {journal} {Physical Review Letters}\ }\textbf {\bibinfo {volume} {127}},\ \bibinfo {pages} {110506} (\bibinfo {year} {2021})}\BibitemShut {NoStop}%
\bibitem [{\citenamefont {Molesky}\ \emph {et~al.}(2022)\citenamefont {Molesky}, \citenamefont {Chao}, \citenamefont {Mohajan}, \citenamefont {Reinhart}, \citenamefont {Chi},\ and\ \citenamefont {Rodriguez}}]{molesky_t_communication_2022}%
  \BibitemOpen
  \bibfield  {author} {\bibinfo {author} {\bibfnamefont {S.}~\bibnamefont {Molesky}}, \bibinfo {author} {\bibfnamefont {P.}~\bibnamefont {Chao}}, \bibinfo {author} {\bibfnamefont {J.}~\bibnamefont {Mohajan}}, \bibinfo {author} {\bibfnamefont {W.}~\bibnamefont {Reinhart}}, \bibinfo {author} {\bibfnamefont {H.}~\bibnamefont {Chi}},\ and\ \bibinfo {author} {\bibfnamefont {A.~W.}\ \bibnamefont {Rodriguez}},\ }\bibfield  {title} {\bibinfo {title} {{$\mathbb{T}$}-operator limits on optical communication: {{Metaoptics}}, computation, and input-output transformations},\ }\href {https://doi.org/10.1103/PhysRevResearch.4.013020} {\bibfield  {journal} {\bibinfo  {journal} {Physical Review Research}\ }\textbf {\bibinfo {volume} {4}},\ \bibinfo {pages} {013020} (\bibinfo {year} {2022})}\BibitemShut {NoStop}%
\bibitem [{\citenamefont {Chao}\ \emph {et~al.}(2022{\natexlab{b}})\citenamefont {Chao}, \citenamefont {Defo}, \citenamefont {Molesky},\ and\ \citenamefont {Rodriguez}}]{chaoMaximumElectromagnetic2022}%
  \BibitemOpen
  \bibfield  {author} {\bibinfo {author} {\bibfnamefont {P.}~\bibnamefont {Chao}}, \bibinfo {author} {\bibfnamefont {R.~K.}\ \bibnamefont {Defo}}, \bibinfo {author} {\bibfnamefont {S.}~\bibnamefont {Molesky}},\ and\ \bibinfo {author} {\bibfnamefont {A.}~\bibnamefont {Rodriguez}},\ }\bibfield  {title} {\bibinfo {title} {Maximum electromagnetic local density of states via material structuring},\ }\bibfield  {journal} {\bibinfo  {journal} {Nanophotonics}\ }\href {https://doi.org/10.1515/nanoph-2022-0600} {10.1515/nanoph-2022-0600} (\bibinfo {year} {2022}{\natexlab{b}})\BibitemShut {NoStop}%
\bibitem [{\citenamefont {Angeris}\ \emph {et~al.}(2023)\citenamefont {Angeris}, \citenamefont {Diamandis}, \citenamefont {Vu{\v c}kovi{\'c}},\ and\ \citenamefont {Boyd}}]{angerisBoundsEfficiency2023}%
  \BibitemOpen
  \bibfield  {author} {\bibinfo {author} {\bibfnamefont {G.}~\bibnamefont {Angeris}}, \bibinfo {author} {\bibfnamefont {T.}~\bibnamefont {Diamandis}}, \bibinfo {author} {\bibfnamefont {J.}~\bibnamefont {Vu{\v c}kovi{\'c}}},\ and\ \bibinfo {author} {\bibfnamefont {S.~P.}\ \bibnamefont {Boyd}},\ }\bibfield  {title} {\bibinfo {title} {Bounds on {{Efficiency Metrics}} in {{Photonics}}},\ }\href {https://doi.org/10.1021/acsphotonics.3c00023} {\bibfield  {journal} {\bibinfo  {journal} {ACS Photonics}\ }\textbf {\bibinfo {volume} {10}},\ \bibinfo {pages} {2521} (\bibinfo {year} {2023})}\BibitemShut {NoStop}%
\bibitem [{\citenamefont {Amaolo}\ \emph {et~al.}(2024{\natexlab{a}})\citenamefont {Amaolo}, \citenamefont {Chao}, \citenamefont {Maldonado}, \citenamefont {Molesky},\ and\ \citenamefont {Rodriguez}}]{amaolo_can_heterostructures_2023}%
  \BibitemOpen
  \bibfield  {author} {\bibinfo {author} {\bibfnamefont {A.}~\bibnamefont {Amaolo}}, \bibinfo {author} {\bibfnamefont {P.}~\bibnamefont {Chao}}, \bibinfo {author} {\bibfnamefont {T.~J.}\ \bibnamefont {Maldonado}}, \bibinfo {author} {\bibfnamefont {S.}~\bibnamefont {Molesky}},\ and\ \bibinfo {author} {\bibfnamefont {A.~W.}\ \bibnamefont {Rodriguez}},\ }\bibfield  {title} {\bibinfo {title} {Can photonic heterostructures provably outperform single-material geometries?},\ }\href@noop {} {\bibfield  {journal} {\bibinfo  {journal} {Nanophotonics}\ }\textbf {\bibinfo {volume} {13}},\ \bibinfo {pages} {283} (\bibinfo {year} {2024}{\natexlab{a}})}\BibitemShut {NoStop}%
\bibitem [{\citenamefont {Mohajan}\ \emph {et~al.}(2023{\natexlab{a}})\citenamefont {Mohajan}, \citenamefont {Chao}, \citenamefont {Jin}, \citenamefont {Molesky},\ and\ \citenamefont {Rodriguez}}]{mohajanFundamentalLimits2023}%
  \BibitemOpen
  \bibfield  {author} {\bibinfo {author} {\bibfnamefont {J.}~\bibnamefont {Mohajan}}, \bibinfo {author} {\bibfnamefont {P.}~\bibnamefont {Chao}}, \bibinfo {author} {\bibfnamefont {W.}~\bibnamefont {Jin}}, \bibinfo {author} {\bibfnamefont {S.}~\bibnamefont {Molesky}},\ and\ \bibinfo {author} {\bibfnamefont {A.~W.}\ \bibnamefont {Rodriguez}},\ }\bibfield  {title} {\bibinfo {title} {Fundamental limits on radiative {$\chi$}{\textsuperscript{(2)}} second harmonic generation},\ }\href {https://doi.org/10.1364/OE.513565} {\bibfield  {journal} {\bibinfo  {journal} {Optics Express}\ }\textbf {\bibinfo {volume} {31}},\ \bibinfo {pages} {44212} (\bibinfo {year} {2023}{\natexlab{a}})}\BibitemShut {NoStop}%
\bibitem [{\citenamefont {Strekha}\ \emph {et~al.}(2024{\natexlab{a}})\citenamefont {Strekha}, \citenamefont {Kr{\"u}ger},\ and\ \citenamefont {Rodriguez}}]{strekhaTraceExpressions2024}%
  \BibitemOpen
  \bibfield  {author} {\bibinfo {author} {\bibfnamefont {B.}~\bibnamefont {Strekha}}, \bibinfo {author} {\bibfnamefont {M.}~\bibnamefont {Kr{\"u}ger}},\ and\ \bibinfo {author} {\bibfnamefont {A.~W.}\ \bibnamefont {Rodriguez}},\ }\bibfield  {title} {\bibinfo {title} {Trace expressions and associated limits for equilibrium {{Casimir}} torque},\ }\href {https://doi.org/10.1103/PhysRevA.109.012813} {\bibfield  {journal} {\bibinfo  {journal} {Physical Review A}\ }\textbf {\bibinfo {volume} {109}},\ \bibinfo {pages} {012813} (\bibinfo {year} {2024}{\natexlab{a}})}\BibitemShut {NoStop}%
\bibitem [{\citenamefont {Amaolo}\ \emph {et~al.}(2024{\natexlab{b}})\citenamefont {Amaolo}, \citenamefont {Chao}, \citenamefont {Maldonado}, \citenamefont {Molesky},\ and\ \citenamefont {Rodriguez}}]{amaoloPhysicalLimits2024}%
  \BibitemOpen
  \bibfield  {author} {\bibinfo {author} {\bibfnamefont {A.}~\bibnamefont {Amaolo}}, \bibinfo {author} {\bibfnamefont {P.}~\bibnamefont {Chao}}, \bibinfo {author} {\bibfnamefont {T.~J.}\ \bibnamefont {Maldonado}}, \bibinfo {author} {\bibfnamefont {S.}~\bibnamefont {Molesky}},\ and\ \bibinfo {author} {\bibfnamefont {A.~W.}\ \bibnamefont {Rodriguez}},\ }\bibfield  {title} {\bibinfo {title} {Physical limits on {{Raman}} scattering: {{The}} critical role of pump and signal co-design},\ }\href {https://doi.org/10.1103/PhysRevA.110.L061501} {\bibfield  {journal} {\bibinfo  {journal} {Physical Review A}\ }\textbf {\bibinfo {volume} {110}},\ \bibinfo {pages} {L061501} (\bibinfo {year} {2024}{\natexlab{b}})}\BibitemShut {NoStop}%
\bibitem [{\citenamefont {Amaolo}\ \emph {et~al.}(2024{\natexlab{c}})\citenamefont {Amaolo}, \citenamefont {Chao}, \citenamefont {Strekha}, \citenamefont {Clarke}, \citenamefont {Mohajan}, \citenamefont {Molesky},\ and\ \citenamefont {Rodriguez}}]{amaoloMaximumShannon2024}%
  \BibitemOpen
  \bibfield  {author} {\bibinfo {author} {\bibfnamefont {A.}~\bibnamefont {Amaolo}}, \bibinfo {author} {\bibfnamefont {P.}~\bibnamefont {Chao}}, \bibinfo {author} {\bibfnamefont {B.}~\bibnamefont {Strekha}}, \bibinfo {author} {\bibfnamefont {S.}~\bibnamefont {Clarke}}, \bibinfo {author} {\bibfnamefont {J.}~\bibnamefont {Mohajan}}, \bibinfo {author} {\bibfnamefont {S.}~\bibnamefont {Molesky}},\ and\ \bibinfo {author} {\bibfnamefont {A.~W.}\ \bibnamefont {Rodriguez}},\ }\href {https://doi.org/10.48550/arXiv.2409.02089} {\bibinfo {title} {Maximum {{Shannon Capacity}} of {{Photonic Structures}}}} (\bibinfo {year} {2024}{\natexlab{c}}),\ \Eprint {https://arxiv.org/abs/2409.02089} {arXiv:2409.02089 [physics]} \BibitemShut {NoStop}%
\bibitem [{\citenamefont {Chao}\ \emph {et~al.}(2025)\citenamefont {Chao}, \citenamefont {Hammond},\ and\ \citenamefont {Johnson}}]{chaoSumofSquaresBounds2025}%
  \BibitemOpen
  \bibfield  {author} {\bibinfo {author} {\bibfnamefont {P.}~\bibnamefont {Chao}}, \bibinfo {author} {\bibfnamefont {I.~M.}\ \bibnamefont {Hammond}},\ and\ \bibinfo {author} {\bibfnamefont {S.~G.}\ \bibnamefont {Johnson}},\ }\href {https://doi.org/10.48550/arXiv.2502.09821} {\bibinfo {title} {Sum-of-{{Squares Bounds}} on {{Surface-Enhanced Raman Scattering}}}} (\bibinfo {year} {2025}),\ \Eprint {https://arxiv.org/abs/2502.09821} {arXiv:2502.09821 [physics]} \BibitemShut {NoStop}%
\bibitem [{\citenamefont {Venkataram}\ \emph {et~al.}(2020)\citenamefont {Venkataram}, \citenamefont {Molesky}, \citenamefont {Jin},\ and\ \citenamefont {Rodriguez}}]{venkataram_fundamental_2020-1}%
  \BibitemOpen
  \bibfield  {author} {\bibinfo {author} {\bibfnamefont {P.~S.}\ \bibnamefont {Venkataram}}, \bibinfo {author} {\bibfnamefont {S.}~\bibnamefont {Molesky}}, \bibinfo {author} {\bibfnamefont {W.}~\bibnamefont {Jin}},\ and\ \bibinfo {author} {\bibfnamefont {A.~W.}\ \bibnamefont {Rodriguez}},\ }\bibfield  {title} {\bibinfo {title} {Fundamental limits to radiative heat transfer: The limited role of nanostructuring in the near-field},\ }\href@noop {} {\bibfield  {journal} {\bibinfo  {journal} {Physical Review Letters}\ }\textbf {\bibinfo {volume} {124}},\ \bibinfo {pages} {013904} (\bibinfo {year} {2020})}\BibitemShut {NoStop}%
\bibitem [{\citenamefont {Li}\ and\ \citenamefont {Hsu}(2022)}]{liThicknessBound2022}%
  \BibitemOpen
  \bibfield  {author} {\bibinfo {author} {\bibfnamefont {S.}~\bibnamefont {Li}}\ and\ \bibinfo {author} {\bibfnamefont {C.~W.}\ \bibnamefont {Hsu}},\ }\bibfield  {title} {\bibinfo {title} {Thickness bound for nonlocal wide-field-of-view metalenses},\ }\href {https://doi.org/10.1038/s41377-022-01038-6} {\bibfield  {journal} {\bibinfo  {journal} {Light: Science \& Applications}\ }\textbf {\bibinfo {volume} {11}},\ \bibinfo {pages} {338} (\bibinfo {year} {2022})}\BibitemShut {NoStop}%
\bibitem [{\citenamefont {Li}\ \emph {et~al.}(2024)\citenamefont {Li}, \citenamefont {Lin},\ and\ \citenamefont {Hsu}}]{liHighefficiencyHighnumericalaperture2024}%
  \BibitemOpen
  \bibfield  {author} {\bibinfo {author} {\bibfnamefont {S.}~\bibnamefont {Li}}, \bibinfo {author} {\bibfnamefont {H.-C.}\ \bibnamefont {Lin}},\ and\ \bibinfo {author} {\bibfnamefont {C.~W.}\ \bibnamefont {Hsu}},\ }\bibfield  {title} {\bibinfo {title} {High-efficiency high-numerical-aperture metalens designed by maximizing the efficiency limit},\ }\href {https://doi.org/10.1364/OPTICA.514907} {\bibfield  {journal} {\bibinfo  {journal} {Optica}\ }\textbf {\bibinfo {volume} {11}},\ \bibinfo {pages} {454} (\bibinfo {year} {2024})}\BibitemShut {NoStop}%
\bibitem [{\citenamefont {Shim}\ \emph {et~al.}(2019)\citenamefont {Shim}, \citenamefont {Fan}, \citenamefont {Johnson},\ and\ \citenamefont {Miller}}]{shim_fundamental_2019}%
  \BibitemOpen
  \bibfield  {author} {\bibinfo {author} {\bibfnamefont {H.}~\bibnamefont {Shim}}, \bibinfo {author} {\bibfnamefont {L.}~\bibnamefont {Fan}}, \bibinfo {author} {\bibfnamefont {S.~G.}\ \bibnamefont {Johnson}},\ and\ \bibinfo {author} {\bibfnamefont {O.~D.}\ \bibnamefont {Miller}},\ }\bibfield  {title} {\bibinfo {title} {Fundamental {{Limits}} to {{Near-Field Optical Response}} over {{Any Bandwidth}}},\ }\href {https://doi.org/10.1103/PhysRevX.9.011043} {\bibfield  {journal} {\bibinfo  {journal} {Physical Review X}\ }\textbf {\bibinfo {volume} {9}},\ \bibinfo {pages} {011043} (\bibinfo {year} {2019})}\BibitemShut {NoStop}%
\bibitem [{\citenamefont {Chao}\ \emph {et~al.}(2022{\natexlab{c}})\citenamefont {Chao}, \citenamefont {Kuate~Defo}, \citenamefont {Molesky},\ and\ \citenamefont {Rodriguez}}]{chao_maximum_2022}%
  \BibitemOpen
  \bibfield  {author} {\bibinfo {author} {\bibfnamefont {P.}~\bibnamefont {Chao}}, \bibinfo {author} {\bibfnamefont {R.}~\bibnamefont {Kuate~Defo}}, \bibinfo {author} {\bibfnamefont {S.}~\bibnamefont {Molesky}},\ and\ \bibinfo {author} {\bibfnamefont {A.}~\bibnamefont {Rodriguez}},\ }\bibfield  {title} {\bibinfo {title} {Maximum electromagnetic local density of states via material structuring},\ }\href@noop {} {\bibfield  {journal} {\bibinfo  {journal} {Nanophotonics}\ }\textbf {\bibinfo {volume} {12}},\ \bibinfo {pages} {549} (\bibinfo {year} {2022}{\natexlab{c}})}\BibitemShut {NoStop}%
\bibitem [{\citenamefont {Armani}\ \emph {et~al.}(2003)\citenamefont {Armani}, \citenamefont {Kippenberg}, \citenamefont {Spillane},\ and\ \citenamefont {Vahala}}]{armani_ultra-high-q_2003}%
  \BibitemOpen
  \bibfield  {author} {\bibinfo {author} {\bibfnamefont {D.~K.}\ \bibnamefont {Armani}}, \bibinfo {author} {\bibfnamefont {T.~J.}\ \bibnamefont {Kippenberg}}, \bibinfo {author} {\bibfnamefont {S.~M.}\ \bibnamefont {Spillane}},\ and\ \bibinfo {author} {\bibfnamefont {K.~J.}\ \bibnamefont {Vahala}},\ }\bibfield  {title} {\bibinfo {title} {Ultra-high-{{Q}} toroid microcavity on a chip},\ }\href {https://doi.org/10.1038/nature01371} {\bibfield  {journal} {\bibinfo  {journal} {Nature}\ }\textbf {\bibinfo {volume} {421}},\ \bibinfo {pages} {925} (\bibinfo {year} {2003})}\BibitemShut {NoStop}%
\bibitem [{\citenamefont {Hsu}\ \emph {et~al.}(2016)\citenamefont {Hsu}, \citenamefont {Zhen}, \citenamefont {Stone}, \citenamefont {Joannopoulos},\ and\ \citenamefont {Solja{\v c}i{\'c}}}]{hsuBoundStates2016}%
  \BibitemOpen
  \bibfield  {author} {\bibinfo {author} {\bibfnamefont {C.~W.}\ \bibnamefont {Hsu}}, \bibinfo {author} {\bibfnamefont {B.}~\bibnamefont {Zhen}}, \bibinfo {author} {\bibfnamefont {A.~D.}\ \bibnamefont {Stone}}, \bibinfo {author} {\bibfnamefont {J.~D.}\ \bibnamefont {Joannopoulos}},\ and\ \bibinfo {author} {\bibfnamefont {M.}~\bibnamefont {Solja{\v c}i{\'c}}},\ }\bibfield  {title} {\bibinfo {title} {Bound states in the continuum},\ }\href {https://doi.org/10.1038/natrevmats.2016.48} {\bibfield  {journal} {\bibinfo  {journal} {Nature Reviews Materials}\ }\textbf {\bibinfo {volume} {1}},\ \bibinfo {pages} {1} (\bibinfo {year} {2016})}\BibitemShut {NoStop}%
\bibitem [{\citenamefont {Johnson}\ \emph {et~al.}(2001)\citenamefont {Johnson}, \citenamefont {Fan}, \citenamefont {Mekis},\ and\ \citenamefont {Joannopoulos}}]{johnsonMultipolecancellationMechanism2001}%
  \BibitemOpen
  \bibfield  {author} {\bibinfo {author} {\bibfnamefont {S.~G.}\ \bibnamefont {Johnson}}, \bibinfo {author} {\bibfnamefont {S.}~\bibnamefont {Fan}}, \bibinfo {author} {\bibfnamefont {A.}~\bibnamefont {Mekis}},\ and\ \bibinfo {author} {\bibfnamefont {J.~D.}\ \bibnamefont {Joannopoulos}},\ }\bibfield  {title} {\bibinfo {title} {Multipole-cancellation mechanism for high-{{Q}} cavities in the absence of a complete photonic band gap},\ }\href {https://doi.org/10.1063/1.1375838} {\bibfield  {journal} {\bibinfo  {journal} {Applied Physics Letters}\ }\textbf {\bibinfo {volume} {78}},\ \bibinfo {pages} {3388} (\bibinfo {year} {2001})}\BibitemShut {NoStop}%
\bibitem [{\citenamefont {Srinivasan}\ and\ \citenamefont {Painter}(2002)}]{srinivasanMomentumSpace2002}%
  \BibitemOpen
  \bibfield  {author} {\bibinfo {author} {\bibfnamefont {K.}~\bibnamefont {Srinivasan}}\ and\ \bibinfo {author} {\bibfnamefont {O.}~\bibnamefont {Painter}},\ }\bibfield  {title} {\bibinfo {title} {Momentum space design of high-{{Q}} photonic crystal optical cavities},\ }\href {https://doi.org/10.1364/OE.10.000670} {\bibfield  {journal} {\bibinfo  {journal} {Optics Express}\ }\textbf {\bibinfo {volume} {10}},\ \bibinfo {pages} {670} (\bibinfo {year} {2002})}\BibitemShut {NoStop}%
\bibitem [{\citenamefont {Maier}\ and\ \citenamefont {Atwater}(2005)}]{maierPlasmonicsLocalization2005}%
  \BibitemOpen
  \bibfield  {author} {\bibinfo {author} {\bibfnamefont {S.~A.}\ \bibnamefont {Maier}}\ and\ \bibinfo {author} {\bibfnamefont {H.~A.}\ \bibnamefont {Atwater}},\ }\bibfield  {title} {\bibinfo {title} {Plasmonics: {{Localization}} and guiding of electromagnetic energy in metal/dielectric structures},\ }\href {https://doi.org/10.1063/1.1951057} {\bibfield  {journal} {\bibinfo  {journal} {Journal of Applied Physics}\ }\textbf {\bibinfo {volume} {98}},\ \bibinfo {pages} {011101} (\bibinfo {year} {2005})}\BibitemShut {NoStop}%
\bibitem [{\citenamefont {Albrechtsen}\ \emph {et~al.}(2022)\citenamefont {Albrechtsen}, \citenamefont {Vosoughi~Lahijani}, \citenamefont {Christiansen}, \citenamefont {Nguyen}, \citenamefont {Casses}, \citenamefont {Hansen}, \citenamefont {Stenger}, \citenamefont {Sigmund}, \citenamefont {Jansen}, \citenamefont {M{\o}rk},\ and\ \citenamefont {Stobbe}}]{albrechtsenNanometerscalePhoton2022}%
  \BibitemOpen
  \bibfield  {author} {\bibinfo {author} {\bibfnamefont {M.}~\bibnamefont {Albrechtsen}}, \bibinfo {author} {\bibfnamefont {B.}~\bibnamefont {Vosoughi~Lahijani}}, \bibinfo {author} {\bibfnamefont {R.~E.}\ \bibnamefont {Christiansen}}, \bibinfo {author} {\bibfnamefont {V.~T.~H.}\ \bibnamefont {Nguyen}}, \bibinfo {author} {\bibfnamefont {L.~N.}\ \bibnamefont {Casses}}, \bibinfo {author} {\bibfnamefont {S.~E.}\ \bibnamefont {Hansen}}, \bibinfo {author} {\bibfnamefont {N.}~\bibnamefont {Stenger}}, \bibinfo {author} {\bibfnamefont {O.}~\bibnamefont {Sigmund}}, \bibinfo {author} {\bibfnamefont {H.}~\bibnamefont {Jansen}}, \bibinfo {author} {\bibfnamefont {J.}~\bibnamefont {M{\o}rk}},\ and\ \bibinfo {author} {\bibfnamefont {S.}~\bibnamefont {Stobbe}},\ }\bibfield  {title} {\bibinfo {title} {Nanometer-scale photon confinement in topology-optimized dielectric cavities},\ }\href {https://doi.org/10.1038/s41467-022-33874-w} {\bibfield  {journal} {\bibinfo  {journal} {Nature Communications}\ }\textbf {\bibinfo {volume}
  {13}},\ \bibinfo {pages} {6281} (\bibinfo {year} {2022})}\BibitemShut {NoStop}%
\bibitem [{\citenamefont {Liang}\ and\ \citenamefont {Johnson}(2013)}]{liangFormulationScalable2013}%
  \BibitemOpen
  \bibfield  {author} {\bibinfo {author} {\bibfnamefont {X.}~\bibnamefont {Liang}}\ and\ \bibinfo {author} {\bibfnamefont {S.~G.}\ \bibnamefont {Johnson}},\ }\bibfield  {title} {\bibinfo {title} {Formulation for scalable optimization of microcavities via the frequency-averaged local density of states},\ }\href@noop {} {\bibfield  {journal} {\bibinfo  {journal} {Optics Express}\ }\textbf {\bibinfo {volume} {21}},\ \bibinfo {pages} {30812} (\bibinfo {year} {2013})}\BibitemShut {NoStop}%
\bibitem [{\citenamefont {Chen}\ \emph {et~al.}(2024)\citenamefont {Chen}, \citenamefont {Christiansen}, \citenamefont {Fan}, \citenamefont {I{\c s}iklar}, \citenamefont {Jiang}, \citenamefont {Johnson}, \citenamefont {Ma}, \citenamefont {Miller}, \citenamefont {Oskooi}, \citenamefont {Schubert}, \citenamefont {Wang}, \citenamefont {Williamson}, \citenamefont {Xue},\ and\ \citenamefont {Zhou}}]{chenValidationCharacterization2024}%
  \BibitemOpen
  \bibfield  {author} {\bibinfo {author} {\bibfnamefont {M.}~\bibnamefont {Chen}}, \bibinfo {author} {\bibfnamefont {R.~E.}\ \bibnamefont {Christiansen}}, \bibinfo {author} {\bibfnamefont {J.~A.}\ \bibnamefont {Fan}}, \bibinfo {author} {\bibfnamefont {G.}~\bibnamefont {I{\c s}iklar}}, \bibinfo {author} {\bibfnamefont {J.}~\bibnamefont {Jiang}}, \bibinfo {author} {\bibfnamefont {S.~G.}\ \bibnamefont {Johnson}}, \bibinfo {author} {\bibfnamefont {W.}~\bibnamefont {Ma}}, \bibinfo {author} {\bibfnamefont {O.~D.}\ \bibnamefont {Miller}}, \bibinfo {author} {\bibfnamefont {A.}~\bibnamefont {Oskooi}}, \bibinfo {author} {\bibfnamefont {M.~F.}\ \bibnamefont {Schubert}}, \bibinfo {author} {\bibfnamefont {F.}~\bibnamefont {Wang}}, \bibinfo {author} {\bibfnamefont {I.~A.~D.}\ \bibnamefont {Williamson}}, \bibinfo {author} {\bibfnamefont {W.}~\bibnamefont {Xue}},\ and\ \bibinfo {author} {\bibfnamefont {Y.}~\bibnamefont {Zhou}},\ }\bibfield  {title} {\bibinfo {title} {Validation and characterization of algorithms and~software
  for photonics inverse design},\ }\href {https://doi.org/10.1364/JOSAB.506412} {\bibfield  {journal} {\bibinfo  {journal} {JOSA B}\ }\textbf {\bibinfo {volume} {41}},\ \bibinfo {pages} {A161} (\bibinfo {year} {2024})}\BibitemShut {NoStop}%
\bibitem [{\citenamefont {Dalklint}\ \emph {et~al.}(2024)\citenamefont {Dalklint}, \citenamefont {Christiansen},\ and\ \citenamefont {Sigmund}}]{dalklintPerformanceBounds2024}%
  \BibitemOpen
  \bibfield  {author} {\bibinfo {author} {\bibfnamefont {A.}~\bibnamefont {Dalklint}}, \bibinfo {author} {\bibfnamefont {R.~E.}\ \bibnamefont {Christiansen}},\ and\ \bibinfo {author} {\bibfnamefont {O.}~\bibnamefont {Sigmund}},\ }\href {https://doi.org/10.48550/arXiv.2410.20375} {\bibinfo {title} {On performance bounds for topology optimization}} (\bibinfo {year} {2024}),\ \Eprint {https://arxiv.org/abs/2410.20375} {arXiv:2410.20375 [cs]} \BibitemShut {NoStop}%
\bibitem [{\citenamefont {Sion}(1958)}]{sionGeneralMinimax1958}%
  \BibitemOpen
  \bibfield  {author} {\bibinfo {author} {\bibfnamefont {M.}~\bibnamefont {Sion}},\ }\bibfield  {title} {\bibinfo {title} {On general minimax theorems},\ }\href {https://doi.org/10.2140/pjm.1958.8.171} {\bibfield  {journal} {\bibinfo  {journal} {Pacific Journal of Mathematics}\ }\textbf {\bibinfo {volume} {8}},\ \bibinfo {pages} {171} (\bibinfo {year} {1958})}\BibitemShut {NoStop}%
\bibitem [{\citenamefont {Molesky}\ \emph {et~al.}()\citenamefont {Molesky}, \citenamefont {Chao}, \citenamefont {Amaolo},\ and\ \citenamefont {Rodriguez}}]{molesky_verlan}%
  \BibitemOpen
  \bibfield  {author} {\bibinfo {author} {\bibfnamefont {S.}~\bibnamefont {Molesky}}, \bibinfo {author} {\bibfnamefont {P.}~\bibnamefont {Chao}}, \bibinfo {author} {\bibfnamefont {A.}~\bibnamefont {Amaolo}},\ and\ \bibinfo {author} {\bibfnamefont {A.~W.}\ \bibnamefont {Rodriguez}},\ }\bibfield  {title} {\bibinfo {title} {Inferring structure via duality for photonic inverse design},\ }\href@noop {} {\bibinfo  {journal} {In preparation}\ }\BibitemShut {NoStop}%
\bibitem [{\citenamefont {Chao}\ \emph {et~al.}(2022{\natexlab{d}})\citenamefont {Chao}, \citenamefont {Strekha}, \citenamefont {Kuate~Defo}, \citenamefont {Molesky},\ and\ \citenamefont {Rodriguez}}]{chao_physical_2022}%
  \BibitemOpen
\bibfield  {journal} {  }\bibfield  {author} {\bibinfo {author} {\bibfnamefont {P.}~\bibnamefont {Chao}}, \bibinfo {author} {\bibfnamefont {B.}~\bibnamefont {Strekha}}, \bibinfo {author} {\bibfnamefont {R.}~\bibnamefont {Kuate~Defo}}, \bibinfo {author} {\bibfnamefont {S.}~\bibnamefont {Molesky}},\ and\ \bibinfo {author} {\bibfnamefont {A.~W.}\ \bibnamefont {Rodriguez}},\ }\bibfield  {title} {\bibinfo {title} {Physical limits in electromagnetism},\ }\href {https://doi.org/10.1038/s42254-022-00468-w} {\bibfield  {journal} {\bibinfo  {journal} {Nature Reviews Physics}\ }\textbf {\bibinfo {volume} {4}},\ \bibinfo {pages} {543} (\bibinfo {year} {2022}{\natexlab{d}})}\BibitemShut {NoStop}%
\bibitem [{\citenamefont {Mohajan}\ \emph {et~al.}(2023{\natexlab{b}})\citenamefont {Mohajan}, \citenamefont {Chao}, \citenamefont {Jin}, \citenamefont {Molesky},\ and\ \citenamefont {Rodriguez}}]{mohajan_fundamental_2023}%
  \BibitemOpen
  \bibfield  {author} {\bibinfo {author} {\bibfnamefont {J.}~\bibnamefont {Mohajan}}, \bibinfo {author} {\bibfnamefont {P.}~\bibnamefont {Chao}}, \bibinfo {author} {\bibfnamefont {W.}~\bibnamefont {Jin}}, \bibinfo {author} {\bibfnamefont {S.}~\bibnamefont {Molesky}},\ and\ \bibinfo {author} {\bibfnamefont {A.~W.}\ \bibnamefont {Rodriguez}},\ }\bibfield  {title} {\bibinfo {title} {Fundamental limits on radiative $\chi$ (2) second harmonic generation},\ }\href@noop {} {\bibfield  {journal} {\bibinfo  {journal} {Optics Express}\ }\textbf {\bibinfo {volume} {31}},\ \bibinfo {pages} {44212} (\bibinfo {year} {2023}{\natexlab{b}})}\BibitemShut {NoStop}%
\bibitem [{\citenamefont {Christiansen}\ and\ \citenamefont {Sigmund}(2021)}]{christiansen_inverse_2021}%
  \BibitemOpen
  \bibfield  {author} {\bibinfo {author} {\bibfnamefont {R.~E.}\ \bibnamefont {Christiansen}}\ and\ \bibinfo {author} {\bibfnamefont {O.}~\bibnamefont {Sigmund}},\ }\bibfield  {title} {\bibinfo {title} {Inverse design in photonics by topology optimization: Tutorial},\ }\href@noop {} {\bibfield  {journal} {\bibinfo  {journal} {JOSA B}\ }\textbf {\bibinfo {volume} {38}},\ \bibinfo {pages} {496} (\bibinfo {year} {2021})}\BibitemShut {NoStop}%
\bibitem [{\citenamefont {Schubert}\ \emph {et~al.}(2022)\citenamefont {Schubert}, \citenamefont {Cheung}, \citenamefont {Williamson}, \citenamefont {Spyra},\ and\ \citenamefont {Alexander}}]{schubert_strict_fabrication_2022}%
  \BibitemOpen
  \bibfield  {author} {\bibinfo {author} {\bibfnamefont {M.~F.}\ \bibnamefont {Schubert}}, \bibinfo {author} {\bibfnamefont {A.~K.~C.}\ \bibnamefont {Cheung}}, \bibinfo {author} {\bibfnamefont {I.~A.~D.}\ \bibnamefont {Williamson}}, \bibinfo {author} {\bibfnamefont {A.}~\bibnamefont {Spyra}},\ and\ \bibinfo {author} {\bibfnamefont {D.~H.}\ \bibnamefont {Alexander}},\ }\bibfield  {title} {\bibinfo {title} {Inverse design of photonic devices with strict foundry fabrication constraints},\ }\href {https://doi.org/10.1021/acsphotonics.2c00313} {\bibfield  {journal} {\bibinfo  {journal} {ACS Photonics}\ }\textbf {\bibinfo {volume} {9}},\ \bibinfo {pages} {2327} (\bibinfo {year} {2022})},\ \Eprint {https://arxiv.org/abs/https://doi.org/10.1021/acsphotonics.2c00313} {https://doi.org/10.1021/acsphotonics.2c00313} \BibitemShut {NoStop}%
\bibitem [{\citenamefont {Boyd}\ and\ \citenamefont {Vandenberghe}(2004)}]{boyd_convex_2004}%
  \BibitemOpen
  \bibfield  {author} {\bibinfo {author} {\bibfnamefont {S.~P.}\ \bibnamefont {Boyd}}\ and\ \bibinfo {author} {\bibfnamefont {L.}~\bibnamefont {Vandenberghe}},\ }\href@noop {} {\emph {\bibinfo {title} {Convex Optimization}}}\ (\bibinfo  {publisher} {{Cambridge University Press}},\ \bibinfo {address} {{Cambridge, UK; New York}},\ \bibinfo {year} {2004})\BibitemShut {NoStop}%
\bibitem [{\citenamefont {Chao}(2023)}]{chaoProbingFundamental2023}%
  \BibitemOpen
  \bibfield  {author} {\bibinfo {author} {\bibfnamefont {P.}~\bibnamefont {Chao}},\ }\emph {\bibinfo {title} {Probing {{Fundamental Performance Limits}} in {{Photonics Design}}}},\ \href@noop {} {Ph.D. thesis},\ \bibinfo  {school} {Princeton University}, \bibinfo {address} {United States -- New Jersey} (\bibinfo {year} {2023})\BibitemShut {NoStop}%
\bibitem [{\citenamefont {Novotny}\ and\ \citenamefont {Hecht}(2012)}]{novotnyPrinciplesNanooptics2012}%
  \BibitemOpen
  \bibfield  {author} {\bibinfo {author} {\bibfnamefont {L.}~\bibnamefont {Novotny}}\ and\ \bibinfo {author} {\bibfnamefont {B.}~\bibnamefont {Hecht}},\ }\href@noop {} {\emph {\bibinfo {title} {Principles of Nano-Optics}}},\ \bibinfo {edition} {second edition}\ ed.\ (\bibinfo  {publisher} {Cambridge University Press},\ \bibinfo {address} {Cambridge},\ \bibinfo {year} {2012})\BibitemShut {NoStop}%
\bibitem [{\citenamefont {Seok}\ \emph {et~al.}(2011)\citenamefont {Seok}, \citenamefont {Jamshidi}, \citenamefont {Kim}, \citenamefont {Dhuey}, \citenamefont {Lakhani}, \citenamefont {Choo}, \citenamefont {Schuck}, \citenamefont {Cabrini}, \citenamefont {Schwartzberg}, \citenamefont {Bokor}, \citenamefont {Yablonovitch},\ and\ \citenamefont {Wu}}]{seokRadiationEngineering2011}%
  \BibitemOpen
  \bibfield  {author} {\bibinfo {author} {\bibfnamefont {T.~J.}\ \bibnamefont {Seok}}, \bibinfo {author} {\bibfnamefont {A.}~\bibnamefont {Jamshidi}}, \bibinfo {author} {\bibfnamefont {M.}~\bibnamefont {Kim}}, \bibinfo {author} {\bibfnamefont {S.}~\bibnamefont {Dhuey}}, \bibinfo {author} {\bibfnamefont {A.}~\bibnamefont {Lakhani}}, \bibinfo {author} {\bibfnamefont {H.}~\bibnamefont {Choo}}, \bibinfo {author} {\bibfnamefont {P.~J.}\ \bibnamefont {Schuck}}, \bibinfo {author} {\bibfnamefont {S.}~\bibnamefont {Cabrini}}, \bibinfo {author} {\bibfnamefont {A.~M.}\ \bibnamefont {Schwartzberg}}, \bibinfo {author} {\bibfnamefont {J.}~\bibnamefont {Bokor}}, \bibinfo {author} {\bibfnamefont {E.}~\bibnamefont {Yablonovitch}},\ and\ \bibinfo {author} {\bibfnamefont {M.~C.}\ \bibnamefont {Wu}},\ }\bibfield  {title} {\bibinfo {title} {Radiation {{Engineering}} of {{Optical Antennas}} for {{Maximum Field Enhancement}}},\ }\href {https://doi.org/10.1021/nl2010862} {\bibfield  {journal} {\bibinfo  {journal} {Nano Letters}\
  }\textbf {\bibinfo {volume} {11}},\ \bibinfo {pages} {2606} (\bibinfo {year} {2011})}\BibitemShut {NoStop}%
\bibitem [{\citenamefont {Wang}\ \emph {et~al.}(2021)\citenamefont {Wang}, \citenamefont {Yu}, \citenamefont {Wang}, \citenamefont {Zhang}, \citenamefont {Kuo}, \citenamefont {Xu},\ and\ \citenamefont {Wang}}]{wangHighQPlasmonic2021}%
  \BibitemOpen
  \bibfield  {author} {\bibinfo {author} {\bibfnamefont {B.}~\bibnamefont {Wang}}, \bibinfo {author} {\bibfnamefont {P.}~\bibnamefont {Yu}}, \bibinfo {author} {\bibfnamefont {W.}~\bibnamefont {Wang}}, \bibinfo {author} {\bibfnamefont {X.}~\bibnamefont {Zhang}}, \bibinfo {author} {\bibfnamefont {H.-C.}\ \bibnamefont {Kuo}}, \bibinfo {author} {\bibfnamefont {H.}~\bibnamefont {Xu}},\ and\ \bibinfo {author} {\bibfnamefont {Z.~M.}\ \bibnamefont {Wang}},\ }\bibfield  {title} {\bibinfo {title} {High-{{Q Plasmonic Resonances}}: {{Fundamentals}} and {{Applications}}},\ }\href {https://doi.org/10.1002/adom.202001520} {\bibfield  {journal} {\bibinfo  {journal} {Advanced Optical Materials}\ }\textbf {\bibinfo {volume} {9}},\ \bibinfo {pages} {2001520} (\bibinfo {year} {2021})}\BibitemShut {NoStop}%
\bibitem [{\citenamefont {Kato}(1995)}]{katoPerturbationTheory1995}%
  \BibitemOpen
  \bibfield  {author} {\bibinfo {author} {\bibfnamefont {T.}~\bibnamefont {Kato}},\ }\href@noop {} {\emph {\bibinfo {title} {Perturbation Theory for Linear Operators}}},\ \bibinfo {edition} {repr. of the 1980 ed}\ ed.,\ Classics in Mathematics\ (\bibinfo  {publisher} {Springer},\ \bibinfo {address} {Berlin Heidelberg},\ \bibinfo {year} {1995})\BibitemShut {NoStop}%
\bibitem [{\citenamefont {Johnson}\ \emph {et~al.}(2002)\citenamefont {Johnson}, \citenamefont {Ibanescu}, \citenamefont {Skorobogatiy}, \citenamefont {Weisberg}, \citenamefont {Joannopoulos},\ and\ \citenamefont {Fink}}]{johnsonPerturbationTheory2002}%
  \BibitemOpen
  \bibfield  {author} {\bibinfo {author} {\bibfnamefont {S.~G.}\ \bibnamefont {Johnson}}, \bibinfo {author} {\bibfnamefont {M.}~\bibnamefont {Ibanescu}}, \bibinfo {author} {\bibfnamefont {M.~A.}\ \bibnamefont {Skorobogatiy}}, \bibinfo {author} {\bibfnamefont {O.}~\bibnamefont {Weisberg}}, \bibinfo {author} {\bibfnamefont {J.~D.}\ \bibnamefont {Joannopoulos}},\ and\ \bibinfo {author} {\bibfnamefont {Y.}~\bibnamefont {Fink}},\ }\bibfield  {title} {\bibinfo {title} {Perturbation theory for {{Maxwell}}'s equations with shifting material boundaries},\ }\href {https://doi.org/10.1103/PhysRevE.65.066611} {\bibfield  {journal} {\bibinfo  {journal} {Physical Review E}\ }\textbf {\bibinfo {volume} {65}},\ \bibinfo {pages} {066611} (\bibinfo {year} {2002})}\BibitemShut {NoStop}%
\bibitem [{\citenamefont {Logan}\ \emph {et~al.}(2024)\citenamefont {Logan}, \citenamefont {Yama},\ and\ \citenamefont {Fu}}]{loganSelectiveActive2024}%
  \BibitemOpen
  \bibfield  {author} {\bibinfo {author} {\bibfnamefont {A.~D.}\ \bibnamefont {Logan}}, \bibinfo {author} {\bibfnamefont {N.~S.}\ \bibnamefont {Yama}},\ and\ \bibinfo {author} {\bibfnamefont {K.-M.~C.}\ \bibnamefont {Fu}},\ }\bibfield  {title} {\bibinfo {title} {Selective active resonance tuning for multi-mode nonlinear photonic cavities},\ }\href {https://doi.org/10.1364/OE.512048} {\bibfield  {journal} {\bibinfo  {journal} {Optics Express}\ }\textbf {\bibinfo {volume} {32}},\ \bibinfo {pages} {13396} (\bibinfo {year} {2024})}\BibitemShut {NoStop}%
\bibitem [{\citenamefont {Wang}\ \emph {et~al.}(2018)\citenamefont {Wang}, \citenamefont {Christiansen}, \citenamefont {Yu}, \citenamefont {M{\o}rk},\ and\ \citenamefont {Sigmund}}]{wang_maximizing_2018}%
  \BibitemOpen
  \bibfield  {author} {\bibinfo {author} {\bibfnamefont {F.}~\bibnamefont {Wang}}, \bibinfo {author} {\bibfnamefont {R.~E.}\ \bibnamefont {Christiansen}}, \bibinfo {author} {\bibfnamefont {Y.}~\bibnamefont {Yu}}, \bibinfo {author} {\bibfnamefont {J.}~\bibnamefont {M{\o}rk}},\ and\ \bibinfo {author} {\bibfnamefont {O.}~\bibnamefont {Sigmund}},\ }\bibfield  {title} {\bibinfo {title} {Maximizing the quality factor to mode volume ratio for ultra-small photonic crystal cavities},\ }\href@noop {} {\bibfield  {journal} {\bibinfo  {journal} {Applied Physics Letters}\ }\textbf {\bibinfo {volume} {113}},\ \bibinfo {pages} {241101} (\bibinfo {year} {2018})}\BibitemShut {NoStop}%
\bibitem [{\citenamefont {Svanberg}(1987)}]{svanbergMethodMoving1987a}%
  \BibitemOpen
  \bibfield  {author} {\bibinfo {author} {\bibfnamefont {K.}~\bibnamefont {Svanberg}},\ }\bibfield  {title} {\bibinfo {title} {The method of moving asymptotes---a new method for structural optimization},\ }\href {https://doi.org/10.1002/nme.1620240207} {\bibfield  {journal} {\bibinfo  {journal} {International Journal for Numerical Methods in Engineering}\ }\textbf {\bibinfo {volume} {24}},\ \bibinfo {pages} {359} (\bibinfo {year} {1987})}\BibitemShut {NoStop}%
\bibitem [{\citenamefont {Johnson}\ \emph {et~al.}(2019)\citenamefont {Johnson} \emph {et~al.}}]{johnson_nlopt_2019}%
  \BibitemOpen
  \bibfield  {author} {\bibinfo {author} {\bibfnamefont {S.~G.}\ \bibnamefont {Johnson}} \emph {et~al.},\ }\href@noop {} {\emph {\bibinfo {title} {The {{NLopt}} Nonlinear Optimization Package (Version 2.6.2)}}}\ (\bibinfo {year} {2019})\BibitemShut {NoStop}%
\bibitem [{\citenamefont {Smith}\ and\ \citenamefont {Pendry}(2006)}]{smithHomogenizationMetamaterials2006}%
  \BibitemOpen
  \bibfield  {author} {\bibinfo {author} {\bibfnamefont {D.~R.}\ \bibnamefont {Smith}}\ and\ \bibinfo {author} {\bibfnamefont {J.~B.}\ \bibnamefont {Pendry}},\ }\bibfield  {title} {\bibinfo {title} {Homogenization of metamaterials by field averaging (invited paper)},\ }\href {https://doi.org/10.1364/JOSAB.23.000391} {\bibfield  {journal} {\bibinfo  {journal} {JOSA B}\ }\textbf {\bibinfo {volume} {23}},\ \bibinfo {pages} {391} (\bibinfo {year} {2006})}\BibitemShut {NoStop}%
\bibitem [{\citenamefont {Tsukerman}(2011)}]{tsukermanEffectiveParameters2011}%
  \BibitemOpen
  \bibfield  {author} {\bibinfo {author} {\bibfnamefont {I.}~\bibnamefont {Tsukerman}},\ }\bibfield  {title} {\bibinfo {title} {Effective parameters of metamaterials: A rigorous homogenization theory via {{Whitney}} interpolation},\ }\href {https://doi.org/10.1364/JOSAB.28.000577} {\bibfield  {journal} {\bibinfo  {journal} {JOSA B}\ }\textbf {\bibinfo {volume} {28}},\ \bibinfo {pages} {577} (\bibinfo {year} {2011})}\BibitemShut {NoStop}%
\bibitem [{\citenamefont {Markel}(2016)}]{markel_maxwell_2016}%
  \BibitemOpen
  \bibfield  {author} {\bibinfo {author} {\bibfnamefont {V.~A.}\ \bibnamefont {Markel}},\ }\bibfield  {title} {\bibinfo {title} {Maxwell {{Garnett}} approximation (advanced topics): Tutorial},\ }\href@noop {} {\bibfield  {journal} {\bibinfo  {journal} {Journal of the Optical Society of America A}\ }\textbf {\bibinfo {volume} {33}},\ \bibinfo {pages} {2237} (\bibinfo {year} {2016})}\BibitemShut {NoStop}%
\bibitem [{\citenamefont {Butcher}\ and\ \citenamefont {High}(2022)}]{butcherAlldielectricMultiresonant2022}%
  \BibitemOpen
  \bibfield  {author} {\bibinfo {author} {\bibfnamefont {A.}~\bibnamefont {Butcher}}\ and\ \bibinfo {author} {\bibfnamefont {A.~A.}\ \bibnamefont {High}},\ }\bibfield  {title} {\bibinfo {title} {All-dielectric multi-resonant bullseye antennas},\ }\href {https://doi.org/10.1364/OE.455232} {\bibfield  {journal} {\bibinfo  {journal} {Optics Express}\ }\textbf {\bibinfo {volume} {30}},\ \bibinfo {pages} {12092} (\bibinfo {year} {2022})}\BibitemShut {NoStop}%
\bibitem [{\citenamefont {Marcatili}(1969)}]{marcatili_bends_1969}%
  \BibitemOpen
  \bibfield  {author} {\bibinfo {author} {\bibfnamefont {E.~a.~J.}\ \bibnamefont {Marcatili}},\ }\bibfield  {title} {\bibinfo {title} {Bends in {{Optical Dielectric Guides}}},\ }\href {https://doi.org/10.1002/j.1538-7305.1969.tb01167.x} {\bibfield  {journal} {\bibinfo  {journal} {Bell System Technical Journal}\ }\textbf {\bibinfo {volume} {48}},\ \bibinfo {pages} {2103} (\bibinfo {year} {1969})}\BibitemShut {NoStop}%
\bibitem [{\citenamefont {Quan}\ and\ \citenamefont {Loncar}(2011)}]{quanDeterministicDesign2011}%
  \BibitemOpen
  \bibfield  {author} {\bibinfo {author} {\bibfnamefont {Q.}~\bibnamefont {Quan}}\ and\ \bibinfo {author} {\bibfnamefont {M.}~\bibnamefont {Loncar}},\ }\bibfield  {title} {\bibinfo {title} {Deterministic design of wavelength scale, ultra-high {{Q}} photonic crystal nanobeam cavities},\ }\href {https://doi.org/10.1364/OE.19.018529} {\bibfield  {journal} {\bibinfo  {journal} {Optics Express}\ }\textbf {\bibinfo {volume} {19}},\ \bibinfo {pages} {18529} (\bibinfo {year} {2011})}\BibitemShut {NoStop}%
\bibitem [{\citenamefont {Seidler}\ \emph {et~al.}(2013)\citenamefont {Seidler}, \citenamefont {Lister}, \citenamefont {Drechsler}, \citenamefont {Hofrichter},\ and\ \citenamefont {St{\"o}ferle}}]{seidlerSlottedPhotonic2013}%
  \BibitemOpen
  \bibfield  {author} {\bibinfo {author} {\bibfnamefont {P.}~\bibnamefont {Seidler}}, \bibinfo {author} {\bibfnamefont {K.}~\bibnamefont {Lister}}, \bibinfo {author} {\bibfnamefont {U.}~\bibnamefont {Drechsler}}, \bibinfo {author} {\bibfnamefont {J.}~\bibnamefont {Hofrichter}},\ and\ \bibinfo {author} {\bibfnamefont {T.}~\bibnamefont {St{\"o}ferle}},\ }\bibfield  {title} {\bibinfo {title} {Slotted photonic crystal nanobeam cavity with an ultrahigh quality factor-to-mode volume ratio},\ }\href {https://doi.org/10.1364/OE.21.032468} {\bibfield  {journal} {\bibinfo  {journal} {Optics Express}\ }\textbf {\bibinfo {volume} {21}},\ \bibinfo {pages} {32468} (\bibinfo {year} {2013})}\BibitemShut {NoStop}%
\bibitem [{\citenamefont {Strekha}\ \emph {et~al.}(2024{\natexlab{b}})\citenamefont {Strekha}, \citenamefont {Chao}, \citenamefont {Defo}, \citenamefont {Molesky},\ and\ \citenamefont {Rodriguez}}]{strekhaSuppressingElectromagnetic2024}%
  \BibitemOpen
  \bibfield  {author} {\bibinfo {author} {\bibfnamefont {B.}~\bibnamefont {Strekha}}, \bibinfo {author} {\bibfnamefont {P.}~\bibnamefont {Chao}}, \bibinfo {author} {\bibfnamefont {R.~K.}\ \bibnamefont {Defo}}, \bibinfo {author} {\bibfnamefont {S.}~\bibnamefont {Molesky}},\ and\ \bibinfo {author} {\bibfnamefont {A.~W.}\ \bibnamefont {Rodriguez}},\ }\bibfield  {title} {\bibinfo {title} {Suppressing electromagnetic local density of states via slow light in lossy quasi-one-dimensional gratings},\ }\href {https://doi.org/10.1103/PhysRevA.109.L041501} {\bibfield  {journal} {\bibinfo  {journal} {Physical Review A}\ }\textbf {\bibinfo {volume} {109}},\ \bibinfo {pages} {L041501} (\bibinfo {year} {2024}{\natexlab{b}})}\BibitemShut {NoStop}%
\bibitem [{\citenamefont {Strekha}\ \emph {et~al.}(2024{\natexlab{c}})\citenamefont {Strekha}, \citenamefont {Chao}, \citenamefont {Defo}, \citenamefont {Molesky},\ and\ \citenamefont {Rodriguez}}]{strekha_suppressing_2024}%
  \BibitemOpen
  \bibfield  {author} {\bibinfo {author} {\bibfnamefont {B.}~\bibnamefont {Strekha}}, \bibinfo {author} {\bibfnamefont {P.}~\bibnamefont {Chao}}, \bibinfo {author} {\bibfnamefont {R.~K.}\ \bibnamefont {Defo}}, \bibinfo {author} {\bibfnamefont {S.}~\bibnamefont {Molesky}},\ and\ \bibinfo {author} {\bibfnamefont {A.~W.}\ \bibnamefont {Rodriguez}},\ }\bibfield  {title} {\bibinfo {title} {Suppressing electromagnetic local density of states via slow light in lossy quasi-one-dimensional gratings},\ }\href@noop {} {\bibfield  {journal} {\bibinfo  {journal} {Physical Review A}\ }\textbf {\bibinfo {volume} {109}},\ \bibinfo {pages} {L041501} (\bibinfo {year} {2024}{\natexlab{c}})}\BibitemShut {NoStop}%
\bibitem [{\citenamefont {Boriskina}\ \emph {et~al.}(2017)\citenamefont {Boriskina}, \citenamefont {Cooper}, \citenamefont {Zeng}, \citenamefont {Ni}, \citenamefont {Tong}, \citenamefont {Tsurimaki}, \citenamefont {Huang}, \citenamefont {Meroueh}, \citenamefont {Mahan},\ and\ \citenamefont {Chen}}]{boriskinaLossesPlasmonics2017}%
  \BibitemOpen
  \bibfield  {author} {\bibinfo {author} {\bibfnamefont {S.~V.}\ \bibnamefont {Boriskina}}, \bibinfo {author} {\bibfnamefont {T.~A.}\ \bibnamefont {Cooper}}, \bibinfo {author} {\bibfnamefont {L.}~\bibnamefont {Zeng}}, \bibinfo {author} {\bibfnamefont {G.}~\bibnamefont {Ni}}, \bibinfo {author} {\bibfnamefont {J.~K.}\ \bibnamefont {Tong}}, \bibinfo {author} {\bibfnamefont {Y.}~\bibnamefont {Tsurimaki}}, \bibinfo {author} {\bibfnamefont {Y.}~\bibnamefont {Huang}}, \bibinfo {author} {\bibfnamefont {L.}~\bibnamefont {Meroueh}}, \bibinfo {author} {\bibfnamefont {G.}~\bibnamefont {Mahan}},\ and\ \bibinfo {author} {\bibfnamefont {G.}~\bibnamefont {Chen}},\ }\bibfield  {title} {\bibinfo {title} {Losses in plasmonics: From mitigating energy dissipation to embracing loss-enabled functionalities},\ }\href {https://doi.org/10.1364/AOP.9.000775} {\bibfield  {journal} {\bibinfo  {journal} {Advances in Optics and Photonics}\ }\textbf {\bibinfo {volume} {9}},\ \bibinfo {pages} {775} (\bibinfo {year} {2017})}\BibitemShut
  {NoStop}%
\bibitem [{\citenamefont {Ahmadi}\ and\ \citenamefont {Majumdar}(2019)}]{ahmadiDSOSSDSOS2019}%
  \BibitemOpen
  \bibfield  {author} {\bibinfo {author} {\bibfnamefont {A.~A.}\ \bibnamefont {Ahmadi}}\ and\ \bibinfo {author} {\bibfnamefont {A.}~\bibnamefont {Majumdar}},\ }\bibfield  {title} {\bibinfo {title} {{{DSOS}} and {{SDSOS Optimization}}: {{More Tractable Alternatives}} to {{Sum}} of {{Squares}} and {{Semidefinite Optimization}}},\ }\href {https://doi.org/10.1137/18M118935X} {\bibfield  {journal} {\bibinfo  {journal} {SIAM Journal on Applied Algebra and Geometry}\ }\textbf {\bibinfo {volume} {3}},\ \bibinfo {pages} {193} (\bibinfo {year} {2019})}\BibitemShut {NoStop}%
\bibitem [{\citenamefont {Wolpert}\ and\ \citenamefont {Macready}(1997)}]{wolpertNoFree1997}%
  \BibitemOpen
  \bibfield  {author} {\bibinfo {author} {\bibfnamefont {D.}~\bibnamefont {Wolpert}}\ and\ \bibinfo {author} {\bibfnamefont {W.}~\bibnamefont {Macready}},\ }\bibfield  {title} {\bibinfo {title} {No free lunch theorems for optimization},\ }\href {https://doi.org/10.1109/4235.585893} {\bibfield  {journal} {\bibinfo  {journal} {IEEE Transactions on Evolutionary Computation}\ }\textbf {\bibinfo {volume} {1}},\ \bibinfo {pages} {67} (\bibinfo {year} {1997})}\BibitemShut {NoStop}%
\bibitem [{\citenamefont {Ho}\ and\ \citenamefont {Pepyne}(2002)}]{hoSimpleExplanation2002}%
  \BibitemOpen
  \bibfield  {author} {\bibinfo {author} {\bibfnamefont {Y.}~\bibnamefont {Ho}}\ and\ \bibinfo {author} {\bibfnamefont {D.}~\bibnamefont {Pepyne}},\ }\bibfield  {title} {\bibinfo {title} {Simple {{Explanation}} of the {{No-Free-Lunch Theorem}} and {{Its Implications}}},\ }\href {https://doi.org/10.1023/A:1021251113462} {\bibfield  {journal} {\bibinfo  {journal} {Journal of Optimization Theory and Applications}\ }\textbf {\bibinfo {volume} {115}},\ \bibinfo {pages} {549} (\bibinfo {year} {2002})}\BibitemShut {NoStop}%
\bibitem [{\citenamefont {Aruliah}\ and\ \citenamefont {Ascher}(2002)}]{aruliahMultigridPreconditioning2002}%
  \BibitemOpen
  \bibfield  {author} {\bibinfo {author} {\bibfnamefont {D.~A.}\ \bibnamefont {Aruliah}}\ and\ \bibinfo {author} {\bibfnamefont {U.~M.}\ \bibnamefont {Ascher}},\ }\bibfield  {title} {\bibinfo {title} {Multigrid {{Preconditioning}} for {{Krylov Methods}} for {{Time-Harmonic Maxwell}}'s {{Equations}} in {{Three Dimensions}}},\ }\href {https://doi.org/10.1137/S1064827501387358} {\bibfield  {journal} {\bibinfo  {journal} {SIAM Journal on Scientific Computing}\ }\textbf {\bibinfo {volume} {24}},\ \bibinfo {pages} {702} (\bibinfo {year} {2002})}\BibitemShut {NoStop}%
\bibitem [{\citenamefont {Vion}\ and\ \citenamefont {Geuzaine}(2018)}]{vionImprovedSweeping2018}%
  \BibitemOpen
  \bibfield  {author} {\bibinfo {author} {\bibfnamefont {A.}~\bibnamefont {Vion}}\ and\ \bibinfo {author} {\bibfnamefont {C.}~\bibnamefont {Geuzaine}},\ }\bibfield  {title} {\bibinfo {title} {Improved sweeping preconditioners for domain decomposition algorithms applied to time-harmonic {{Helmholtz}} and {{Maxwell}} problems},\ }\href {https://doi.org/10.1051/proc/201861093} {\bibfield  {journal} {\bibinfo  {journal} {ESAIM: Proceedings and Surveys}\ }\textbf {\bibinfo {volume} {61}},\ \bibinfo {pages} {93} (\bibinfo {year} {2018})}\BibitemShut {NoStop}%
\bibitem [{\citenamefont {Bonazzoli}\ \emph {et~al.}(2019)\citenamefont {Bonazzoli}, \citenamefont {Dolean}, \citenamefont {Graham}, \citenamefont {Spence},\ and\ \citenamefont {Tournier}}]{bonazzoliDomainDecomposition2019}%
  \BibitemOpen
  \bibfield  {author} {\bibinfo {author} {\bibfnamefont {M.}~\bibnamefont {Bonazzoli}}, \bibinfo {author} {\bibfnamefont {V.}~\bibnamefont {Dolean}}, \bibinfo {author} {\bibfnamefont {I.}~\bibnamefont {Graham}}, \bibinfo {author} {\bibfnamefont {E.}~\bibnamefont {Spence}},\ and\ \bibinfo {author} {\bibfnamefont {P.-H.}\ \bibnamefont {Tournier}},\ }\bibfield  {title} {\bibinfo {title} {Domain decomposition preconditioning for the high-frequency time-harmonic {{Maxwell}} equations with absorption},\ }\href {https://doi.org/10.1090/mcom/3447} {\bibfield  {journal} {\bibinfo  {journal} {Mathematics of Computation}\ }\textbf {\bibinfo {volume} {88}},\ \bibinfo {pages} {2559} (\bibinfo {year} {2019})}\BibitemShut {NoStop}%
\bibitem [{\citenamefont {Zhang}\ and\ \citenamefont {Zhang}(2018)}]{zhangFFTbasedAlgorithm2018}%
  \BibitemOpen
  \bibfield  {author} {\bibinfo {author} {\bibfnamefont {B.}~\bibnamefont {Zhang}}\ and\ \bibinfo {author} {\bibfnamefont {R.}~\bibnamefont {Zhang}},\ }\bibfield  {title} {\bibinfo {title} {An {{FFT-based Algorithm}} for {{Efficient Computation}} of {{Green}}'s {{Functions}} for the {{Helmholtz}} and {{Maxwell}}'s {{Equations}} in {{Periodic Domains}}},\ }\href {https://doi.org/10.1137/18M1165621} {\bibfield  {journal} {\bibinfo  {journal} {SIAM Journal on Scientific Computing}\ }\textbf {\bibinfo {volume} {40}},\ \bibinfo {pages} {B915} (\bibinfo {year} {2018})}\BibitemShut {NoStop}%
\bibitem [{\citenamefont {Harris}\ \emph {et~al.}(2019)\citenamefont {Harris}, \citenamefont {Langston}, \citenamefont {L{\'e}tourneau}, \citenamefont {Papanicolaou}, \citenamefont {Ezick},\ and\ \citenamefont {Lethin}}]{harrisFastLargeScale2019}%
  \BibitemOpen
  \bibfield  {author} {\bibinfo {author} {\bibfnamefont {M.~T.}\ \bibnamefont {Harris}}, \bibinfo {author} {\bibfnamefont {M.~H.}\ \bibnamefont {Langston}}, \bibinfo {author} {\bibfnamefont {P.-D.}\ \bibnamefont {L{\'e}tourneau}}, \bibinfo {author} {\bibfnamefont {G.}~\bibnamefont {Papanicolaou}}, \bibinfo {author} {\bibfnamefont {J.}~\bibnamefont {Ezick}},\ and\ \bibinfo {author} {\bibfnamefont {R.}~\bibnamefont {Lethin}},\ }\bibfield  {title} {\bibinfo {title} {Fast {{Large-Scale Algorithm}} for {{Electromagnetic Wave Propagation}} in {{3D Media}}},\ }in\ \href {https://doi.org/10.1109/HPEC.2019.8916219} {\emph {\bibinfo {booktitle} {2019 {{IEEE High Performance Extreme Computing Conference}} ({{HPEC}})}}}\ (\bibinfo {year} {2019})\ pp.\ \bibinfo {pages} {1--7}\BibitemShut {NoStop}%
\bibitem [{\citenamefont {Lazarov}\ and\ \citenamefont {Sigmund}(2011)}]{lazarovFiltersTopology2011}%
  \BibitemOpen
  \bibfield  {author} {\bibinfo {author} {\bibfnamefont {B.~S.}\ \bibnamefont {Lazarov}}\ and\ \bibinfo {author} {\bibfnamefont {O.}~\bibnamefont {Sigmund}},\ }\bibfield  {title} {\bibinfo {title} {Filters in topology optimization based on {{Helmholtz-type}} differential equations},\ }\href {https://doi.org/10.1002/nme.3072} {\bibfield  {journal} {\bibinfo  {journal} {International Journal for Numerical Methods in Engineering}\ }\textbf {\bibinfo {volume} {86}},\ \bibinfo {pages} {765} (\bibinfo {year} {2011})}\BibitemShut {NoStop}%
\bibitem [{\citenamefont {Kuster}\ \emph {et~al.}(2025)\citenamefont {Kuster}, \citenamefont {Augenstein}, \citenamefont {Hern{\'a}ndez}, \citenamefont {Rockstuhl},\ and\ \citenamefont {Sturges}}]{kusterInverseDesign2025}%
  \BibitemOpen
  \bibfield  {author} {\bibinfo {author} {\bibfnamefont {O.}~\bibnamefont {Kuster}}, \bibinfo {author} {\bibfnamefont {Y.}~\bibnamefont {Augenstein}}, \bibinfo {author} {\bibfnamefont {R.~N.}\ \bibnamefont {Hern{\'a}ndez}}, \bibinfo {author} {\bibfnamefont {C.}~\bibnamefont {Rockstuhl}},\ and\ \bibinfo {author} {\bibfnamefont {T.~J.}\ \bibnamefont {Sturges}},\ }\href {https://doi.org/10.48550/arXiv.2501.05900} {\bibinfo {title} {Inverse {{Design}} of {{3D Nanophotonic Devices}} with {{Structural Integrity Using Auxiliary Thermal Solvers}}}} (\bibinfo {year} {2025}),\ \Eprint {https://arxiv.org/abs/2501.05900} {arXiv:2501.05900 [physics]} \BibitemShut {NoStop}%
\end{thebibliography}%

\newpage

\pagebreak
\appendix


\section{Numerical Evaluation of the Dual Bound}

\subsection{Finite Difference Discretization of the QCQP}

This Appendix details the numerical approach we took to calculate dual bounds given by \eqref{eq:QCQP} and \eqref{eq:dualopt}. We stress that the structure optimization \eqref{eq:topopt}, field optimization \eqref{eq:QCQP}, and dual optimization \eqref{eq:dualopt} are all well defined mathematical problems independent of any discretization scheme used to numerically solve them. In prior work, we have computed bounds for 3D spherical design domains using a spectral basis based on vector spherical harmonics~\cite{molesky_global_2020,molesky_hierarchical_2020} and bounds for a 2D infinite half-space using Fourier and Laplace Transforms~\cite{chaoMaximumElectromagnetic2022}. Different numerical results under different discretizations for the same objective and constraints in \eqref{eq:QCQP} is a numerical convergence issue in approaching the true value of the dual bound. 

In this work, we used a finite-difference frequency domain (FDFD) discretization for both TO and dual bounds calculations. FDFD is conceptually simple and easy to adapt to various problems with general geometries. Under FDFD, vector fields are represented by finite vectors over a discrete grid, and integral operators are represented by matrices. Given the polarization field $\bm P$ is restricted to the design domain $\Omega$, to numerically evaluate the dual bounds we only need to consider the vector fields and Green's function over $\Omega$. We shall denote the FDFD discretizations of vector fields using bold lower case letters and operators using bold upper case letters. The discretized \eqref{eq:QCQP} in the main text thus reads

\begin{subequations}
\label{eq:disc_QCQP}
\begin{align}
     \max_{\vp} \quad &f(\vp) \equiv \Re{\vo^\dagger \vp} - \vp^\dagger \mO \vp \label{eq:quad_obj} \\
     \text{s.t. } \forall \Omega_j \quad &\Re{\vei^\dagger \mI_j \vp - \vp^\dagger  \bm U \mI_j \vp} = 0 \label{eq:re_cstrt}\\
    & \Re{\vei^\dagger (i \mI_j) \vp - \vp^\dagger (i \bm U \mI_j) \vp} = 0 \label{eq:im_cstrt}
\end{align}
\end{subequations}
where \eqref{eq:re_cstrt} and \eqref{eq:im_cstrt} are the discretized real and imaginary parts of a single constraint in \eqref{eq:QCQP}, the integration over $\Omega_j$ in \eqref{eq:QCQP} is represented by a diagonal spatial projection matrix $\mI_j$, and we have defined $\mU \equiv \chi^{-1\dagger} - \mGv^\dagger$ for notational convenience. The vector $\vo$ and Hermitian matrix $\mO$ represent the general form of a quadratic objective function $f(\vp)$; for the LDOS example of the main text, $\mO = 0$ and $\vo = -\frac{1}{2} \mGv^\dagger \bm j$ where $\bm j$ is a discretized dipole source. 

The Lagrangian of this QCQP is given by 
\begin{subequations}
\begin{equation}
    \Lag(\vphi_R, \vphi_I, \vp) = \Re{\vz(\vphi_R,\vphi_I)^\dagger \vp} - \vp^\dagger \mZ(\vphi_R,\vphi_I) \vp
\end{equation}
where $\vphi_R$, $\vphi_I$ are vectors of Lagrange multipliers for the constraints \eqref{eq:re_cstrt} and \eqref{eq:im_cstrt}, respectively, and we have defined
\begin{align}
    \vz(\vphi_R,\vphi_I) &= \vo + \sum_{j} (\phi_{R,j} - i\phi_{I,j}) \mI_j \vei, \\
    \mZ(\vphi_R,\vphi_I) &= \mO + \mU \sum_{j} (\phi_{R,j} + i\phi_{I,j}) \mI_j. 
\end{align}
\end{subequations}

The dual function over the multipliers is $\dual(\vphi_R, \vphi_I) = \max_{\vp} \Lag(\vphi_R,\vphi_I,\vp)$; this maximum is finite if and only if $\mZ \succeq 0$, in which case the optimal (dual) polarization current is $\vp_{\dual}(\vphi) = \mZ(\vphi)^{-1} \vz(\vphi)$. Strictly speaking we should use the pseudo-inverse to account for the possibility that $\mZ$ may be singular, but if $\vz$ has overlap with nullspace of $\mZ$ the dual is also infinite so we will abuse notation somewhat. The dual problem to arrive at the tightest bound is thus given by
\begin{align*}
    \min_{\vphi} \quad &\dual(\vphi) = \vz(\vphi)^\dagger \mZ(\vphi)^{-1} \vz(\vphi) \numthis \\
    &\mZ(\vphi) \succeq 0
\end{align*}

A straightforward application of the chain rule shows that the dual derivative for any given multiplier is the matching constraint violation $C_j(\vp_{\dual}(\vphi))$:
\begin{equation}
    \pdv{\dual}{\phi_{R,j}} = \Re{\vei^\dagger \mI_j \vp_{\dual}(\vphi) - \vp_{\dual}(\vphi)^\dagger  \bm U \mI_j \vp_{\dual}(\vphi)}
\end{equation}
and similar for $\pdv{\dual}{\phi_{I,j}}$. The Hessian of $\dual$ can also be derived by further differentiation. 

\subsection{Sparse Formulation} \label{asec:sparse}
\begin{figure}
    \centering
    \includegraphics[width=\linewidth]{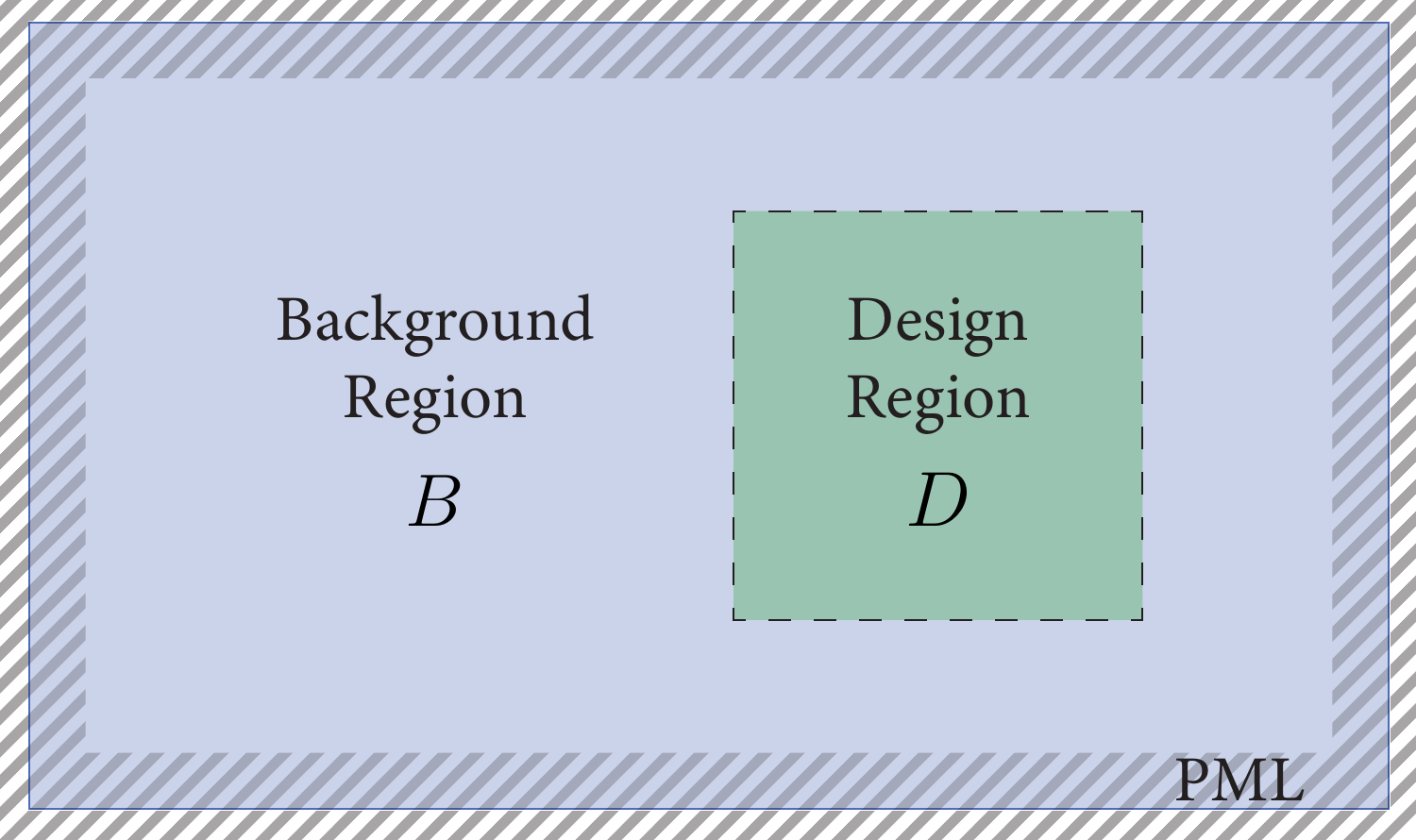}
    \caption{Schematic showing the design and background regions surrounded by a perfrectly matched layer (PML) region that enforces Sommerfield radiation conditions, ensuring the Maxwell operator \(\mathcal M\) is invertible.}
    \label{fig:region_schematic}
\end{figure}
The vacuum Green's function is the inverse of the sparse operator \(\mathcal M =\nabla\times\nabla\times-\omega^2 \mI \), and is thus dense in general. 
In block matrix form, it takes the form
\begin{equation} \label{eq:fullGvac}
    \vb{G} = \begin{bmatrix}
        \bm{G}_{0,DD} & \bm{G}_{0,DB} \\ \bm{G}_{0,BD} & \bm{G}_{0,BB}
    \end{bmatrix}
\end{equation}
where \(D\) and \(B\) represent the design and background regions, respectively (Fig.~\ref{fig:region_schematic}), and \(\bm{G}_{0,XY}\) maps currents in \(Y\) to fields in \(X\). 
Since the polarization is only non-zero in the design region, \(\mGv \equiv \bm{G}_{0,DD}\) in~\eqref{eq:disc_QCQP} and \(\mU\) in general. While \(\mGv\) is used in other sections to denote this operator, in this section we will explicitly specify source and field regions. 
Similarly, \(\vei\) is a vector represented the incident field projected into the design region. 

The dense nature of \(\bm{G}_{0,DD}\) means that the matrix \(\mZ(\vphi)\) is also a dense matrix, making solving for the optimal \(\vp_\dual\), and thus evaluating the dual value and gradient, numerically expensive and memory intensive. 
To greatly improve the scaling of this method, and thus allow for a much greater number of projection constraints \(\mI_j\), we may transform the problem to leverage the sparsity of the Maxwell operator \(\mathcal M\). 
This operator may also be split into its background and design sub-blocks, giving 
\begin{equation}
    \vb G = \omega^2 \begin{bmatrix}
        \mathcal M_{BB} & \mathcal M_{BD} \\ \mathcal M_{DB} & \mathcal M_{DD}
    \end{bmatrix}^{-1}.
\end{equation}
Leveraging block inversion and the definition \eqref{eq:fullGvac}, we find that 
\begin{equation}
    \bm{G}_{0,DD}^{-1} = \dfrac{1}{\omega^2} \left( \mathcal M_{DD} - \mathcal M_{DB} \mathcal M_{BB}^{-1} \mathcal M_{BD} \right),
\end{equation}
which is a sparse matrix in a localized basis representation (e.g., the finite difference pixel basis). 
\(\mathcal M_{BB}^{-1} \mathcal M_{BD}\) may be computed via a matrix linear solve \(\mathcal M_{BB} \bm X = \mathcal M_{BD}\). 

Now, we may place identity matrices \(\bm{G}^{-1}_{0,DD} \bm{G}_{0,DD}\) to re-write~\eqref{eq:disc_QCQP} as
\begin{equation}
\begin{aligned}
     \max_{\vp} \quad& \Re{\vo^\dagger \bm{G}^{-1}_{0,DD} \bm{G}_{0,DD} \vp}  \\ & -\vp^\dagger \bm{G}^{\dagger}_{0,DD} \bm{G}^{-\dagger}_{0,DD} \mO \bm{G}^{-1}_{0,DD} \bm{G}_{0,DD} \vp  \\
     \text{s.t. } \forall \Omega_j \quad &\Re \big\{ \vei^\dagger \mI_j \bm{G}^{-1}_{0,DD} \bm{G}_{0,DD} \vp \\ & - \vp^\dagger  \bm{G}^{\dagger}_{0,DD} \bm{G}^{-\dagger}_{0,DD} \bm U \mI_j \bm{G}^{-1}_{0,DD} \bm{G}_{0,DD} \vp \big\} = 0 \\
    & \Re \big\{ \vei^\dagger (i \mI_j) \bm{G}^{-1}_{0,DD} \bm{G}_{0,DD} \vp \\ & - \vp^\dagger \bm{G}^{\dagger}_{0,DD} \bm{G}^{-\dagger}_{0,DD} (i \bm U \mI_j)  \bm{G}^{-1}_{0,DD} \bm{G}_{0,DD} \vp \big\} = 0.
\end{aligned}
\end{equation}
Now, taking \(\bm{G}_{0,DD} \vp \equiv \vp'\) to be the new optimization variable, and defining a new sparse operator \(\bm{G}^{-\dagger}_{0,DD} \left( \chi^{-\dagger} - \bm{G}^{\dagger}_{0,DD} \right) \mI_j \bm{G}^{-1}_{0,DD} \equiv \mU' \mI_j \), we arrive at the new problem 
\begin{equation}
\label{eq:sparse_QCQP}
\begin{aligned}
     \max_{\vp'} \quad &f(\vp') \equiv \Re{\vo^\dagger \bm{G}^{-1}_{0,DD} \vp'} \\ &\hspace{3em} - \vp'^\dagger\bm{G}^{-\dagger}_{0,DD} \mO \bm{G}^{-1}_{0,DD} \vp' \\
     \text{s.t. } \forall \Omega_j \quad &\Re{\vei^\dagger \mI_j \bm{G}^{-1}_{0,DD} \vp' - \vp'^\dagger  \bm U' \mI_j \vp'} = 0 \\
    & \Re{\vei^\dagger (i \mI_j) \bm{G}^{-1}_{0,DD} \vp' - \vp'^\dagger (i \bm U' \mI_j) \vp'} = 0 
\end{aligned}
\end{equation}
which is mathematically equivalent to~\eqref{eq:disc_QCQP} under a change of variables. 
We similarly define \(\vz'\),  \(\mZ'\), \(\vo'\), and \(\mO'\).
Now, all constraints and objective matrices are sparse, meaning that \(\mZ'\) is sparse, making solving for \(\vp_{\dual}\) and the gradients of the dual function computationally tractable for larger system sizes. 
For an estimate on the sparsity of these operators, we note that for a system size \(4\lambda \times 4\lambda\) at \(40\) pixels per wavelength resolution, \(\bm{G}^{-1}_{0,DD}\) has a density of \(0.08\%\), and \(\mZ'\) of \(0.2\%\). 
This is used throughout this paper and our prior work to greatly speed up bound calculations.  

\section{Efficient Calculation of Dual Solution: Generalized Constraint Descent} \label{asec:gcd}

To get the tightest dual bounds from the QCQP formulation, we would like to include as many constraints as possible: for an FDFD discretization with $N$ pixels in the design region, there is a total of $N$ complex pixel level constraints. 
While the dual problem is convex and we can compute the dual derivatives, in practice for low loss $\chi$ the dual problem may be ill-conditioned. This leads to the computational challenge that first order methods converge very slowly, but second order methods are impractical due to large $N$. In this section, we describe in detail a heuristic of approaching the pixel-level dual optimum without ever having to simultaneously deal with a large number of primal QCQP constraints. This is achieved by a nested dual optimization process where the inner optimization finds the dual optimum of a QCQP with a small ($\lesssim 10$) number of intelligently selected constraints, and the outer optimization modifies the QCQP constraints to obtain a new QCQP with a smaller dual optimum. By keeping the number of constraints small at all times, this allows for the use of second order optimization methods in the inner optimization. 

\subsection{Generalized projection constraints and Lagrange multiplier subspaces}
Suppose we impose pixel level constraints in \eqref{eq:disc_QCQP}: this means that the diagonal projection matrices $\mI_j$ are ``one-hot'': $\mI_{j,kl} = \delta_{kj}\delta_{lj}$. We can thus re-write the pixel contraint dual as
\begin{align*}
    &\dual(\vphi_R,\vphi_I) = \max_{\vp} \quad f(\vp) + \\
    & \Re{\vei^\dagger \sum_{j=1}^N (\phi_{R,j} + i\phi_{I,j}) \mI_j \vp - \vp^\dagger \mU \sum_{j=1}^N (\phi_{R,j} + i\phi_{I,j}) \mI_j \vp } \\
    &= \max_{\vp} \quad f(\vp) + \Re{\vei^\dagger \Diag(\vphi)\vp - \vp^\dagger \mU \Diag(\vphi) \vp}.
\end{align*}
where we have defined the complex multiplier vector $\vphi = \vphi_R + i\vphi_I$, and $\Diag(\vphi)$ is a diagonal matrix with $\vphi$ on its diagonal.
The dual bound is then given by
\begin{equation}
    \mathcal{B}(\mathbb{C}^N) \equiv \min_{\vphi \in \mathbb{C}^N} \quad \dual(\vphi)
\end{equation}
where we implicitly define $\dual(\vphi) = \infty$ when $\mZ(\vphi) \nsucceq 0$. We highlight the fact that the dual optimization is over the vector space $\mathbb{C}^N$; using this notation we can denote looser bounds that are obtained via imposing fewer constraints as
\begin{equation}
    \mathcal{B}(\V) = \min_{\vphi \in \V} \max_{\vp} \quad f(\vp) +  \Re{\vei^\dagger \Diag(\vphi)\vp - \vp^\dagger \mU \Diag(\vphi) \vp}
    \label{eq:dualB_V}
\end{equation}
where $\V$ is a linear subspace of $\mathbb{C}^N$. Suppose that a basis of $\V$ is $\{\vv_1,\cdots,\vv_m\}$: a QCQP that gives the dual bound in \eqref{eq:dualB_V} is
\begin{subequations}
\label{eq:gen_QCQP}
\begin{align*}
     &\max_{\vp} \quad f(\vp) \numthis \\
     &\text{s.t. } \forall k \in \{1,\cdots,m\} \\
     &\Re{\vei^\dagger \Diag(\vv_k) \vp - \vp^\dagger  \bm U \Diag(\vv_k) \vp} = 0 \numthis \label{eq:gen_cstrt}
\end{align*}
\end{subequations}
with a corresponding Lagrangian
\begin{equation} \begin{aligned}
    \Lag_{\V}(\vgma,\vp) = f(\vp) + \Re\bigg\{ &\vei^\dagger \Diag\left(\sum_{k=1}^m \vgma_k \vv_k \right) \vp \\ 
    &- \vp^\dagger \mU \Diag\left(\sum_{k=1}^m \vgma_k \vv_k \right) \vp \bigg\}
\end{aligned} \end{equation}
and dual function $\dual_{\V}(\vgma)$:
\begin{align*}
    \dual_{\V}(\vgma \in \mathbb{R}^m) = \dual(\vphi \in \V) = \max_{\vp} \; \Lag_{\V}(\vgma, \vp).
\end{align*}
Note that the basis $\{\vv_k\}$ is over the real field $\mathbb{R}$: while the vectors $\vv_k$ have complex entries, their linear combinations are taken with real coefficients, i.e., the Lagrangian multipliers $\vgma_k$. Because $\vv_k$ are general vectors in $\mathbb{C}_n$, $\Diag(\vv_k)$ is no longer a simple projection onto some (discretized) spatial region, but a generalized, ``weighted'' projection. Under this framework, the QCQP with pixel level constraints corresponds to (\ref{eq:gen_QCQP}) with $\{\vv_k\}$ the canonical basis of $\mathbb{C}^N$ over $\mathbb{R}$. Given a different basis to $\mathbb{C}^N$, we can form an equivalent primal QCQP with the same dual bound $\mathcal{B}(\mathbb{C}^N)$. The process of tightening the dual bound through the addition of primal constraints can be understood as gradually increasing the dimension of $\V \subset \mathbb{C}^N$; the sections below detail a heuristic for doing such dual subspace expansion. 

\subsection{Heuristic for multiplier subspace expansion}
Suppose we have evaluated $\mathbb{B}(\V)$ with $\V = \text{span}(\vv_1, \cdots, \vv_m)$. Denote the optimal multipliers as $\vgma_{\dual} = \arg \max_{\vgma} \dual_{\V}(\vgma)$, and the corresponding optimal primal vector os $\vp_{\dual} = \arg \max_{\vp} \Lag_{\V}(\vgma_{\dual}, \vp)$. 
We now wish to introduce an additional basis vector $\vv_{m+1}$ to obtain a tighter bound $\mathbb{B}(\widetilde{\V})$ where $\widetilde{\V} = \V \oplus \text{span}\{\vv_{m+1}\}$. 
From the perspective of the updated dual optimization $\max_{\tilde{\vgma} \in \mathbb{R}^{m+1}} \dual_{\widetilde{\V}} (\tilde{\vgma})$, this amounts to initializing at the Lagrange multiplier vector $\tilde{\vgma}_0$ with $\tilde{\vgma}_{0,k} = \vgma_{\dual,k}$ for $1 \leq k \leq m$ and $\tilde{\vgma}_{0,m+1} = 0$. Our goal is thus to select $\vv_{m+1}$ such that the additional dual degree of freedom $\tilde{\vgma}_{m+1}$ is useful for further reduction of the bound. 

\subsubsection{Maximizing constraint violation}
A natural idea is to choose $\vv_{m+1}$ such that $\pdv{\dual_{\widetilde{\V}}}{\tilde{\vgma}_{m+1}} \big|_{\tilde{\vgma} = \tilde{\vgma_0}}$ is large. For any given $\vv_{m+1}$, the dual multiplier derivative is just the value of the primal constraint violation:
\begin{align*}
    \pdv{\dual}{\tilde{\vgma}_{m+1}} \bigg|_{\tilde{\vgma} = \tilde{\vgma_0}}
    &= \Re \bigg\{ \vei^\dagger \Diag(\vv_{m+1}) \vp_{\dual} \\
    & \hspace{3em} - \vp_{\dual}^\dagger \mU\Diag(\vv_{m+1}) \vp_{\dual} \bigg\} \\
    & = \Re \bigg\{ \sum_{j=1}^N ( \bm e_{i,j}^* \vv_{m+1,j} \vp_{\dual,j} \\
    & \hspace{3em} - (\mU^\dagger \vp_{\dual})^*_j \vv_{m+1,j} \vp_{\dual,j} ) \bigg\} \\
    & = \sum_{j=1}^N \Re{[\bm e_{i,j}^* \vp_{\dual,j} - (\mU^\dagger \vp_{\dual})^*_j \vp_{\dual,j}] \vv_{m+1,j}}.   \numthis \label{eq:dualgrad_sumform}
\end{align*}

Each component of $\vv_{m+1}$ thus contributes independently to the dual derivative: assuming for now that each component has magnitude 1, the optimal choice is then
\begin{align*}
    \vv_{m+1,j} &= e^{i \alpha_j}, \\
    \alpha_j &= \pi - \arg( \bm e_{i,j}^* \vp_{\dual,j} - (\mU^\dagger \vp_{\dual})^*_j \vp_{\dual,j} ). \numthis \label{eq:optcstrt_angle}
\end{align*}
This makes all terms within the sum of \eqref{eq:dualgrad_sumform} real and negative, aligning the complex phase of their contributions. Since we are minimizing the dual function, the direction of descent for the new multiplier $\tilde{\vgma}_{m+1}$ would be for it to become more positive. 

\subsubsection{Avoiding semi-definite boundary}
While a larger constraint violation implies a larger dual derivative, this dual derivative cannot be fully exploited if increasing $\tilde{\vgma}_{m+1}$ from 0 immediately runs into the semi-definite boundary $\mZ \succeq 0$. To avoid this, we can also try to choose $\vv_{m+1}$ such that increasing $\tilde{\vgma}_{m+1}$ from 0 moves away from the semi-definite boundary. Define the smallest eigenvalue of $\mZ(\tilde{\vgma})$ as $\eta$ and the corresponding normalized eigenvector as $\vx$: $\mZ(\tilde{\vgma}) \vx = \eta \vx$. What we want to acieve is $\pdv{\eta}{\tilde{\vgma}_{m+1}}\big|_{\tilde{\vgma}=\tilde{\vgma}_0} > 0$. By the Feynman-Hellmann theorem, we have
\begin{align*}
    \pdv{\eta}{\tilde{\vgma}_{m+1}}\big|_{\tilde{\vgma}=\tilde{\vgma}_0} &= \Re\{\vx^\dagger \mU \Diag(\vv_{m+1}) \vx \} \\
    &= \sum_{j=1}^N \Re{(\mU^\dagger \vx)^*_j \vx_j \vv_{m+1,j}}. \numthis \label{eq:mineiggrad_sumform}
\end{align*}
Similar to \eqref{eq:dualgrad_sumform}, each component of $\vv_{m+1}$ contributes independently: assuming for now hat $|\vv_{m+1,j}| = 1$, the optimal choice for maximizing $\pdv{\eta}{\tilde{\vgma}_{m+1}}$ is then
\begin{align*}
    \vv_{m+1,j} &= e^{i \beta_j} \\
    \beta_j &= -\arg((\mU^\dagger \vx)^*_j \vx_j). \numthis \label{eq:opteig_angle}
\end{align*}

\subsubsection{Combining ideas}
How can we simultaneously take care of maximizing new constraint violation and avoiding the semi-definite boundary? While (\ref{eq:optcstrt_angle}) and (\ref{eq:opteig_angle}) give the optimal angles for the complex phasors of $\vv_{m+1,j}$ to maximize dual and minimum eigenvalue gradients, respectively, as long as the argument of $\vv_{m+1,j}$ falls within $\pi/2$ of the optimal angles of each, the contributions from each term in the sums of (\ref{eq:dualgrad_sumform}) and (\ref{eq:mineiggrad_sumform}) would still aligned with the same sign after taking the real part. A simple way to find a compromise direction is then to define
\begin{equation}
    \vv_{m+1,j} = e^{i\alpha_j} + e^{i\beta_j}. \label{eq:compromise_angle}
\end{equation}
which has an argument within $\pi/2$ of both optimal angles $\alpha_j$ and $\beta_j$.  

Note that this specifies the phase angle of each $\vv_{m+1,j}$; the relative magnitude of each $\vv_{m+1,j}$ is as yet unspecified. We currently scale $|\vv_{m+1,j}|$ proportional to the constraint violation contribution $|\bm{e}_{i,j}^* \vp_{\dual,j} - (\mU^\dagger \vp_{\dual})_j^* \vp_{\dual,j}|$, though numerical experiments suggest that keeping all $|\vv_{m+1,j}|$ equal also works well in practice. 

Before incorporating $\vv_{m+1,j}$ into the constraint basis $\{\vv_k\}$, it is a good idea to orthonormalize $\vv_{m+1,j}$ via Gram-Schmidt process against the existing constraint basis vectors; this avoids ill-conditioning of the dual problem due to nearly linearly dependent constraints. 

In principle, starting with just a subspace spanned by a single constraint vector $\vv_1$, we can keep on adding new basis vectors; when the total number of basis vectors is $2N$, then we will have converged upon the dual bound. In practice when $N$ is very big, this may not be computationally feasible: to limit the total number of constraints we ever have to deal with at a time, we can combine adding new constraint basis vectors with merging existing constraint basis vectors. 

\subsection{Dual space contraction}
For a given constraint subspace $\V = \text{span}(\vv_1, \cdots, \vv_m)$ with optimal Lagrange multiplier vector $\vgma_{\dual} = \arg \max_{\vgma} \dual_{\V}(\vgma)$, we can define a contracted 1D constraint subspace $\V' = \text{span}(\vv')$ where $\vv' = \sum_{k=1}^m \vgma_{\dual,k} \vv_k$. It is clear that $\V' \subset \V$ and $\mathcal{B}(\V) = \mathcal{B}(\V')$. We can then restart from $\V'$ and add new constraint vectors; the dual bound remains the same under this contraction of the dual space and will continue to improve with the addition of new vectors. 

\subsection{Generalized Constraint Descent} \label{asec:gcd_final}
\begin{figure}[h!]
    \centering
    \includegraphics[width=\linewidth]{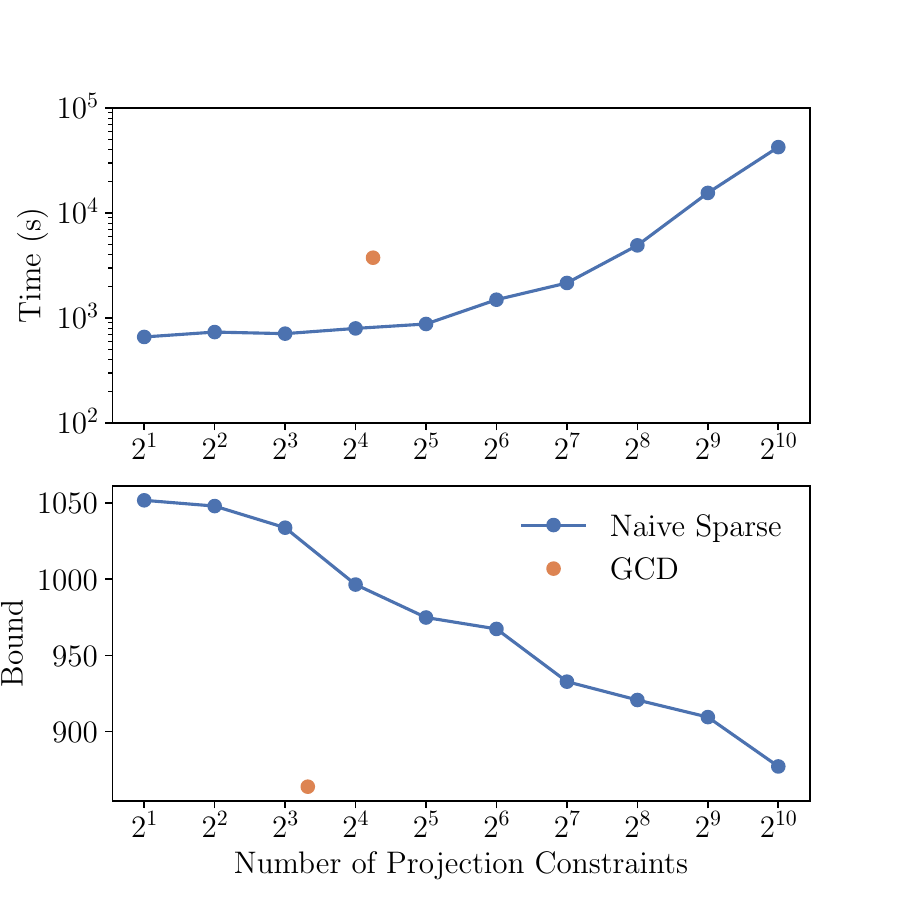}
    \caption{Computation time and limit value as a function of number of projection constraints, for the \(L=4\lambda\) case in Fig.~\ref{fig:TOresults}, for naively calculating a limit leveraging the sparse formulation in Appendix~\ref{asec:sparse} and the gradient coordinate descent procedure in Appendix~\ref{asec:gcd_final}. Results show that for a modest number of projection constraints (five), GCD can achieve a bound better than that given by \(2^{10}\) projection constraints in approximately one-tenth the time. }
    \label{fig:gcd_timing}
\end{figure}

Combining the ideas from the previous sections, we have a generalized constraint descent (GCD) for approaching the pixel level dual bound $\mathbb{B}(\mathbb{C}^N)$: start with an initial small constraint subspace $\V \subset \mathbb{C}^N$. Expand $\V$ with new constraint vectors using the heuristics of maximizing constraint violation and avoiding the semi-definite boundary. When $\text{dim}(\V)$ goes beyond a certain threshold, contract $\V$ back down to a 1D subspace (or more generally, any small number of dimensions) and repeat. With each changing of the subspace the dual bound is guaranteed to be non-increasing. 
Fig.~\ref{fig:gcd_timing} shows how the computation time and bound scale as a function of projection constraints for the GCD algorithm vs. enforcing all constraints in the sparse formulation. 
To formulate a dual problem with \(n\) constraints, we form \(I_j\) by partitioning an identity matrix of dimension \(N\) into \(n\) disjoint diagonal blocks of size \(\frac{N}{n}\times \frac{N}{n}\), setting all other entries to zero.
Data shows that GCD with ten constraints can achieve a bound lower than the naive sparse method with \(2^{10}\) constraints in \(8\%\) of the time for the case studied in Fig.~\ref{fig:TOresults} with \(L=4\lambda\) and a resolution of \(40\) pixels per wavelength. 
The benefit of using GCD is typically greatest for larger system sizes where pixel-level constraints are computationally intractable. 
In Fig.\ref{fig:TOresults} of the main text, ``local dual" bounds and verlan design were obtained using the generalized constraint descent algorithm detailed above with up to 10 changing generalized constraints until the dual bound converged. 


\end{document}